\documentclass[12pt]{article}
\usepackage[reqno]{amsmath}
\usepackage{epsfig}
\usepackage{array}
\usepackage{float}
\usepackage{bbm}
\usepackage{a4}
\usepackage{epsfig}
\usepackage{a4wide}
\usepackage{wasysym}
\def\be{\begin{equation}}
\def\ee{\end{equation}}
\def\gs{\mathrel{
   \rlap{\raise 0.511ex \hbox{$>$}}{\lower 0.511ex \hbox{$\sim$}}}}
\def\ls{\mathrel{
   \rlap{\raise 0.511ex \hbox{$<$}}{\lower 0.511ex \hbox{$\sim$}}}}

\newcommand{\ba}{\begin{array}{c}}
\newcommand{\baz}{\begin{array}{cc}}
\newcommand{\bad}{\begin{array}{ccc}}
\newcommand{\bav}{\begin{array}{cccc}}
\newcommand{\bea}{\begin{equation} \begin{array}{c}}
\newcommand{\eea}{ \end{array} \end{equation}}
\newcommand{\ea}{\end{array}}
\newcommand{\D}{\displaystyle}
\newcommand{\dms}{\mbox{$\Delta m^2_{\odot}$}}
\newcommand{\dma}{\mbox{$\Delta m^2_{\rm A}$}}
\newcommand{\meff}{\mbox{$\left| < \! m \! > \right|$}}

\def\gtap{\mathrel{
   \rlap{\raise 0.511ex \hbox{$>$}}{\lower 0.511ex \hbox{$\sim$}}}}
\def\ltap{\mathrel{
   \rlap{\raise 0.511ex \hbox{$<$}}{\lower 0.511ex \hbox{$\sim$}}}}

\newcommand{\betabeta}{\mbox{$(\beta \beta)_{0 \nu}  $}}

\newcommand{\pmns}{\mbox{$ U_{\rm PMNS}$}}
\newcommand{\lam}{\mbox{$ U_{\lambda}$}}
\newcommand{\bima}{\mbox{$ U_{\rm bimax}$}}
\newcommand{\ts}{\mbox{$ \tan^2 \theta_{\rm sol}$}}
\newcommand{\sa}{\mbox{$ \sin^2 2\theta_{\rm atm}$}}

\begin{document}

\title{
\vspace{-2cm}
\hfill {\small SISSA 7/2004/EP}\\
\vspace{-0.3cm}
\hfill {\small hep-ph/0401206} \\ 
\vskip 0.8cm
\bf On Deviations from Bimaximal Neutrino Mixing
}
% \end{center}
\author{
P.H.~Frampton$^a$, 
$\;$ S.T.~Petcov$^b$\thanks{Also at: Institute of 
Nuclear Research and Nuclear Energy,
Bulgarian Academy of Sciences, 1784 Sofia, Bulgaria}~, 
$\;$ and W.~Rodejohann$^b$\\ \\
{ \normalsize \it $^a$Department of Physics and Astronomy}\\ 
{  \normalsize \it University of North Carolina, 
Chapel Hill, NC 27599-3255, USA}\\ \\
{\normalsize \it $^b$Scuola Internazionale Superiore di Studi Avanzati}\\
{\normalsize \it Via Beirut 2--4, I-34014 Trieste, Italy}\\
{\normalsize and}\\
{\normalsize \it Istituto Nazionale di Fisica Nucleare}\\
{\normalsize \it Sezione di Trieste, I-34014 Trieste, Italy}
}
\date{}
\maketitle
\thispagestyle{empty}
\vspace{-0.8cm}
\begin{abstract}
\noindent 

 The 
% 3x3  unitary 
PMNS neutrino mixing matrix 
$U_{\rm PMNS}$ is in general a product 
of two unitary matrices
$U_{\rm lep}$ and $U_{\nu}$
arising from the diagonalization of 
the charged lepton and neutrino mass matrices,
$U_{\rm PMNS} = U^{\dagger}_{\rm lep} U_{\nu}$.
Assuming that  $U_{\nu}$ is a bimaximal mixing matrix,
we investigate the possible forms of $U_{\rm lep}$.
We identify three possible
generic structures of $U_{\rm lep}$, 
which are compatible with 
the existing data on neutrino mixing.
One corresponds to a hierarchical
``CKM--like'' matrix. In this case 
relatively large values of 
the solar neutrino mixing angle
$\theta_{\rm sol}$,  
and of $|U_{e3}|^2 \equiv |(U_{\rm PMNS})_{e3}|^2$,
are typically predicted, 
$\tan^2\theta_{\rm sol} \gtap 0.42$,
$|U_{e3}|^2 \gtap 0.02$,
while the atmospheric
neutrino mixing angle $\theta_{\rm atm}$ can
deviate noticeably from $\pi/4$,
$\sin^22\theta_{\rm atm} \gtap 0.95$. 
The second corresponds to 
one of the mixing angles in $U_{\rm lep}$
being equal to $\pi/2$, 
and predicts practically maximal atmospheric 
neutrino mixing
$\sin^2 2 \theta_{\rm atm} \simeq 1$.
Large atmospheric neutrino 
mixing, $\sin^22\theta_{\rm atm} \gtap 0.95$,
is naturally predicted by the third possible 
generic structure of $U_{\rm lep}$, 
which corresponds to all three
mixing angles in $U_{\rm lep}$ being large.
We focus especially on the case of 
CP--nonconservation, analyzing it in detail. 
We show how the CP--violating 
phases, arising from the diagonalization 
of the neutrino and charged lepton mass 
matrices, contribute to the measured  
neutrino mixing observables. 

\end{abstract}

\newpage
\section{\label{sec:intro}Introduction}
\vspace{-0.2cm}
\hskip 0.8cm  After the spectacular results obtained in the 
experimental studies of neutrino oscillations in the 
last two and a half years or so 
(see, e.g., \cite{PPVenice03,BCGPRSNO3} for a summary), 
understanding the detailed structure and the origin of 
neutrino masses and mixing is of prime
importance. Progress in understanding
the origin of the Pontecorvo--Maki--Nakagawa--Sakata
(PMNS) mixing matrix in the weak charged lepton current
\cite{BPont57} can lead to a complete solution of the 
fundamental problem regarding the structure
of the neutrino mass spectrum, which can be with normal 
or inverted hierarchy, or of quasi--degenerate
type. It can also help gain significant insight on,
or even answer the fundamental question of, 
CP--violation in the lepton sector.

  The fact that, according to the existing data, the
atmospheric neutrino mixing angle is maximal, or close 
to maximal \cite{SKatm03}, $\theta_{\rm atm} \sim \pi/4$, the 
solar neutrino mixing angle is relatively large
\cite{SNO3,BCGPRSNO3},
$\theta_{\odot} \sim \pi/5.4$ 
($\sin^2\theta_{\odot} \cong 0.30$), and that the 
mixing angle $\theta$ limited by the CHOOZ and Palo
Verde experiments \cite{CHOOZ,PaloV} is small, 
$\sin^2\theta < 0.074$ (99.73\% C.L.) \cite{BCGPRSNO3},
suggests that the PMNS neutrino mixing matrix
\footnote{Throughout this article 
the case of 
three flavour neutrino mixing is considered.},
$U_{\rm PMNS}$, can originate from a 3$\times$3 unitary
mixing matrix having a bimaximal mixing form,
$U_{\rm bimax}$. In the case of CP--invariance 
in the lepton sector one has
%%%%%%%%%%%%%%%%%%%%%%%%%%%%%%%%%%%%%
\be
\label{eq:Ubimax}
U_{\rm bimax} = 
\left(
\bad  
\frac{1}{\sqrt{2}} &  \frac{1}{\sqrt{2}} &  0 \\[0.3cm]
-\frac{1}{2} &  \frac{1}{2} &  \frac{1}{\sqrt{2}} \\[0.3cm]
\frac{1}{2} &  -\frac{1}{2} &  \frac{1}{\sqrt{2}} \\[0.3cm]
\ea 
\right)~.
\ee
%%%%%%%%%%%%%%%%%%%%%%%%%%%%%%%%%%%
%
\noindent  The assumption of exact equality
$U_{\rm PMNS} = U_{\rm bimax}$ implies
$\theta_{\odot} = \pi/4$, which
is ruled out at more than 5 s.d.\ by
the existing solar neutrino data
\cite{SNO3,BCGPRSNO3}. However,
the deviation of $\theta_{\odot}$ from $\pi/4$
can be described by a relatively small
mixing parameter \cite{WR0309}: $\lambda = 
\sin (\pi/4 -\pi/5.4) \cong 0.20$. Thus, 
the neutrino mixing matrix can have the form
%%%%%%%%%%%%%%%%%%%%%%%%
\be 
\pmns = U_\lambda^\dagger \, U_{\rm bimax}~.
\label{Ulam}
\ee
%%%%%%%%%%%%%%%%%%%%%%%%%%%%
%
\noindent It is natural to suppose that
$U_\lambda^\dagger$ and  $U_{\rm bimax}$
in Eq.\ (\ref{Ulam}) arise from
the diagonalization 
of the charged lepton and neutrino mass
matrices, respectively.

   The existing data show also that
\cite{BCGPRSNO3,SKatm03} 
$\dms \ll |\dma|$,
where $\dms > 0$ and  $\dma$
are the neutrino mass squared differences 
driving the solar and atmospheric neutrino
oscillations. This fact and the preceding
considerations suggest further that
$U_{\rm bimax}$ can arise from the 
diagonalization of a neutrino mass term of 
Majorana type having a specific 
symmetry. In the absence of charged lepton
mixing ($U_\lambda = {\mathbbm 1}$)
this symmetry could correspond 
to the conservation of the {\it non--standard} 
lepton charge \cite{STPPD82} $L' = L_e - L_{\mu} - L_{\tau}$,
where $L_l$, $l=e,\mu,\tau$, are the 
electron, muon and 
tauon lepton charges. The indicated symmetry 
cannot be exact: it has to be broken mildly
by the neutrino mass term in order to ensure
that $\Delta m^2_{\odot} \neq 0$, and by
the charged lepton mass term 
($U_\lambda \neq {\mathbbm 1}$),
and/or by the neutrino mass term, 
in order to guarantee
that  $\theta_{\odot} \neq \pi/4$.
 
   The possibility expressed by Eq.\ (\ref{Ulam})
and described above has been discussed
in the context of grand unified theories
by many authors (see, e.g., \cite{GUTs,steve}). 
It has also been investigated 
phenomenologically first 
a long time ago (in a different context) 
in \cite{STPPD82} and more recently 
in \cite{Xing,GTani1,GTani2} (see also \cite{jezsum}). 
Assuming that Eq.\ (\ref{Ulam}) holds,
we perform in the present article
a systematic study of the possible
forms of the matrix $U_\lambda^\dagger$ 
in  Eq.\ (\ref{Ulam}),
which are compatible with the existing data 
on neutrino mixing and oscillations.
The case of CP--nonconservation
is of primary interest and 
is analyzed in detail.
We show, in particular, 
how the CP--violating phases, 
arising from the diagonalization
of the neutrino and charged lepton mass 
matrices, can influence the form 
of $U_\lambda$ and can
contribute in the measured CP--conserving
and CP--violating observables, 
related to neutrino mixing.
The analysis presented here 
can be considered as a
continuation of the earlier studies 
quoted, e.g., in \cite{GUTs,steve,Xing,GTani1,GTani2}.
However, it overlaps little with them.
\vspace{-0.4cm}
%%%%%%%%%%%%%%%%%%%%%%%%%%%%%%%%%%%%%%%

\section{\label{sec:intro2} The Deviations from Bimaximal Mixing}

%%%%%%%%%%%%%%%%%%%%%%%%%%%%%%%%%%%%%%
\vspace{-0.2cm}
\hskip 0.8cm We will employ in what follows the 
standard parametrization of the PMNS matrix:
%%%%%%%%%%%%%%%%%%%%%%%%%%%%
\bea \label{eq:Upara}
\pmns = \left( \bad 
c_{12} c_{13} & s_{12} c_{13} & s_{13}  \\[0.2cm] 
-s_{12} c_{23} - c_{12} s_{23} s_{13} e^{i \delta} 
& c_{12} c_{23} - s_{12} s_{23} s_{13} e^{i \delta} 
& s_{23} c_{13} e^{i \delta} \\[0.2cm] 
s_{12} s_{23} - c_{12} c_{23} s_{13} e^{i \delta} & 
- c_{12} s_{23} - s_{12} c_{23} s_{13} e^{i \delta} 
& c_{23} c_{13} e^{i \delta} \\ 
               \ea   \right) 
 {\rm diag}(1, e^{i \alpha}, e^{i \beta}) \, , 
\eea
%%%%%%%%%%%%%%%%%%%%%
%
\noindent where we have used the usual
notations $c_{ij} = \cos\theta_{ij}$, 
$s_{ij} = \sin\theta_{ij}$,
$\delta$ is the Dirac CP--violation phase, 
$\alpha$ and $\beta$ are two possible Majorana 
CP--violation phases \cite{BHP80,Doi81}.
If we identify 
the two independent neutrino mass squared 
differences in this case,
$\Delta m^2_{21}$ and $\Delta m^2_{31}$,  
with the neutrino mass squared differences
which induce the solar and atmospheric 
neutrino oscillations, $\dms = \Delta m^2_{21} > 0$,
$\dma = \Delta m^2_{31}$,
one has: $\theta_{12} = \theta_{\rm sol}$, 
$\theta_{23} = \theta_{\rm atm}$, 
and $\theta_{13} = \theta$. 
The ranges of values of
the three neutrino 
mixing angles, which are allowed 
at 1 s.d.\ (3 s.d.) by the current 
solar and atmospheric neutrino data and by the data
from the reactor antineutrino experiments
CHOOZ and KamLAND, read \cite{BCGPRSNO3,bari}:
%%%%%%%%%%%%%%%%%%%%%%%%%%%%%%%%%%%%%%%%%%%%%%
\be 
\label{eq:range}
\ba
0.35~(0.27) \leq \tan^2 \theta_{\rm sol} \equiv 
\tan^2 \theta_{12} \leq 0.52~(0.72)~,\\[0.3cm]
|U_{e3}|^2 = \sin^2\theta_{13} <  0.029~(0.074)~, \\[0.2cm]
\sin^2 2 \theta_{\rm atm} \equiv \sin^2 2 \theta_{23} \geq 0.95~(0.85)~.
\ea
\ee
%%%%%%%%%%%%%%%%%%%%%%%%%%%%%%%%%%%%%%%%%%%
%
  As is well--known, the oscillations 
between flavour neutrinos
depend on the Dirac phase $\delta$, but 
are insensitive to the Majorana CP--violating phases 
$\alpha$ and $\beta$ \cite{BHP80,Lang86}.
Information about these phases can be obtained,
in principle, in neutrinoless double beta decay 
(\betabeta--decay) experiments \cite{BGKP96,BPP1,0vbbCP}.

  Let us define the matrix $U_\lambda$ 
through Eq.\ (\ref{Ulam}),
$\pmns \equiv U_\lambda^\dagger \, U_{\rm bimax}$,
where the matrix $U_{\rm bimax}$ is given by Eq.\ (\ref{eq:Ubimax}).
The latter would have coincided with the 
PMNS matrix if the neutrino mixing were exactly
bimaximal. The bimaximal neutrino mixing 
can be obtained, e.g., by exploiting the
flavor neutrino symmetry corresponding 
to the conservation of the non--standard
lepton charge  $L' = (L_e - L_\mu - L_\tau)$
\cite{STPPD82} (see also \cite{CLSP83,BiPet87}). 
The deviations of the neutrino mixing from the
exact bimaximal mixing form can be described,
as was shown in \cite{WR0309}, 
by a real parameter $\lambda \sim 0.2$, 
which was introduced in the 
following ``flexible'' parametrization
of three elements of $\pmns$:  
%%%%%%%%%%%%%%%%%%%%%%%%%%%%%%%%%%%%%
\bea 
\label{eq:master}
U_{e2} = \sqrt{\frac{1}{2}} \, (1 - \lambda)~,~~~
U_{e3} = A \, \lambda^n~,~~~
U_{\mu 3} = \sqrt{\frac{1}{2}} \, (1 - B \, \lambda^m) \, e^{i \delta} ~,
\eea
%%%%%%%%%%%%%%%%%%%%%%%%%%%%%%%%%%%
%
\noindent where $A$ and $B$ are real parameters
of order one.  The integer numbers $m$ and $n$ 
can be chosen according to the
improved limits on, or precise values of,
$\sin^2\theta_{13}$ and $\sin^2 2 \theta_{23}$. 
The fact that solar neutrino mixing is 
now confirmed (at more than 5 s.d.) to be non--maximal 
means that $\lambda \neq 0$ and 
the best--fit value of $\ts = 0.43$ \cite{BCGPRSNO3}   
corresponds to $\lambda \simeq 0.23$. 

% \noindent 
   Taking the purely phenomenological parametrization 
(\ref{eq:master}) at face value and expressing  
the PMNS matrix in terms of $\lambda$, $A$ and $B$, one can solve 
Eq.\ (\ref{Ulam}) for \lam{} in order to obtain 
the physical ranges of values of the
parameters involved.
It was found in \cite{WR0309} that 
the structure of \lam{} can at leading order be 
the unit matrix plus corrections of order $\lambda$. 

   In the present article we perform a systematic study
of the possible forms of \lam{} allowed by the current data.
The neutrino mixing phenomenology can 
be described, for example, by 
\lam{} having a ``CKM--like'' structure, as 
has been discussed in \cite{GTani1,GTani2,WR0309}. 
Here we show, in particular,
that the restriction 
to a hierarchical form for 
\lam{} is not necessary. 
Counter--intuitive forms of \lam{},
and/or the presence of 
CP--violating phases, can produce 
naturally the observed deviation from 
$\pi/4$ of $\theta_{\rm sol}$ as well as 
the requisite smallness of  
$|U_{e3}|$. 
\vspace{-0.4cm}
%%%%%%%%%%%%%%%%%%%%%%%%%%%%%%%

\section{\label{sec:CPC} The Case of CP Conservation}

%%%%%%%%%%%%%%%%%%%%%%%%%%%%%%%%%
\vspace{-0.2cm}
\hskip 0.8cm  In this case 
\lam{} is an orthogonal matrix. 
We will use the standard parametrization 
(see Eq.\ (\ref{eq:Upara})) for \lam{}. 
Let us denote the angles in \lam{} as
$\theta'_{12}$, $\theta'_{23}$, and $\theta'_{13}$,
and define $\sin\theta'_{ij} \equiv \lambda_{ij}$.
We shall treat $\lambda_{12}$, $\lambda_{23}$
and  $\lambda_{13}$ as free parameters
without assuming any hierarchy relation 
between them. In the case of CP--conservation
under discussion one can limit the analysis
to the case $0 \leq \theta'_{ij} \leq \pi$,
and correspondingly to $0 \leq \lambda_{ij} \leq 1$.
The results for $\lambda_{ij} <0$
can be obtained formally from those derived
for $\lambda_{ij} > 0$ by making the
change i) $\lambda_{23} \rightarrow -\lambda_{23}$,
and/or ii) $\lambda_{13} \rightarrow -\lambda_{13}$
as long as one keeps $\lambda_{12} > 0$.
The change $\lambda_{12} \rightarrow -\lambda_{12}$
should be done simultaneously with the change
$\lambda_{13} \rightarrow -\lambda_{13}$ and in this case
only the ``solutions'' with $|\lambda_{13}| > |\lambda_{12}|$
should be considered. In the latter case both signs of
$\lambda_{23}$ are possible.

    Using Eqs.\ (\ref{Ulam}) and (\ref{eq:Ubimax}), 
we have determined the regions of values of
$\lambda_{12}$, $\lambda_{23}$
and  $\lambda_{13}$ for which 
the values of the 
neutrino mixing angles
$\theta_{12} = \theta_{\rm sol}$, 
$\theta_{23} = \theta_{\rm atm}$, 
and $\theta_{13} = \theta$, lying
within their 1$\sigma$ and 3$\sigma$ 
ranges, Eq.\ (\ref{eq:range}), can 
be reproduced. The results are 
shown graphically in Fig.\ \ref{fig:mostgencpc}.
One can clearly identify three different cases:
%%%%%%%%%%%%%%%%%%%%%%%%%%%%%%%%%%%%
\begin{itemize}
\item[(i)] all $\lambda_{ij} \ls 0.35$ (``small''),   
\item[(ii)] $\lambda_{23} = + 1$ and $\lambda_{12, 13} \ls 0.35$ (``small''),  
\item[(iii)] all $\lambda_{ij} \gs 0.40$ (``large''). 
\end{itemize}
%%%%%%%%%%%%%%%%%%%%%%%%%%%%
We shall discuss next these three cases in detail. 
The resulting structures 
of \lam{} will turn out to be rather 
different in the three cases. 

\vspace{-0.4cm}
%%%%%%%%%%%%%%%%%%%%%%%%%%%%%%%%%%%%%%%%%%%%%%%%

\subsection{\label{sec:hie} Small $\lambda_{ij}$}

%%%%%%%%%%%%%%%%%%%%%%%%%%%%%%%%%%%%%%%%%%
\vspace{-0.2cm}
\hskip 0.8cm Multiplying $\lam\!\!^\dagger$ by \bima{} 
and expressing $\tan^2\theta_{\rm sol}$, 
$|U_{e3}|^2$ and $\sin^2 2 \theta_{\rm atm}$
in terms of the three small parameters 
$\lambda_{ij}$, one finds: 

%%%%%%%%%%%%%%%%%%%%%%%%%%%%%%%%%%%%%%%%%%%%%%
\be \label{eq:mixcpc}
\ba 
\tan^2 \theta_{\rm sol} = \frac{\D |U_{e1}|^2}{\D |U_{e2}|^2} \simeq 
1 - 2 \sqrt{2} \, (\lambda_{12} - \lambda_{13}) + 4 \, 
(\lambda_{12} - \lambda_{13} )^2 - 2 \sqrt{2} \, \lambda_{23} \, 
(\lambda_{12} + \lambda_{13}) + {\cal{O}}(\lambda^3)~, \\[0.3cm]
|U_{e3}| \simeq \left| \frac{\D \lambda_{12} + \lambda_{13}}
{\D \sqrt{2}} -
\frac{\D \lambda_{23}}{\D \sqrt{2}} \, (\lambda_{12} - \lambda_{13} ) 
+ {\cal{O}}(\lambda^3) \right|~, \\ %[0.3cm]
\sin^2 2 \theta_{\rm atm} \equiv 
\frac{\D 4 \, |U_{\mu 3}|^2 \, |U_{\tau 3}|^2}{\D (1 - |U_{e3}|^2)^2}  
\simeq 1 - 4 \, \lambda_{23}^2  %-  (\lambda_{12} + \lambda_{13})^2 
+ {\cal{O}}(\lambda^3)~.
\ea 
\ee
%%%%%%%%%%%%%%%%%%%%%%%%%%%%%%%%%%%%%%%%%%%%%%%%
%
\noindent The key quantities are therefore 
$\lambda_{23}$ and the sum and the difference 
of $\lambda_{12}$ and $\lambda_{13}$. 
To leading order in the $\lambda$--parameters, 
the deviation from maximal 
solar neutrino mixing 
and zero $U_{e3}$ are proportional to $\lambda$,  
while the atmospheric neutrino mixing
is close to maximal with only 
quadratic corrections to $(1 - \sin^2 2 \theta_{\rm atm})$. 
The requirement of $\sin^2 2 \theta_{\rm atm} \geq 0.85~(0.95)$ 
leads to order $\lambda^2$ to the restriction 
$\lambda_{23} \ls 0.19~(0.11)$.  

% \noindent 
Assuming that the oscillation parameters 
lie within their 1$\sigma$ allowed ranges and 
neglecting terms of order $\lambda^2$,
one finds from Eq.\ (\ref{eq:mixcpc})
that the term $(\lambda_{12} - \lambda_{13})$ 
has to be positive and rather large 
($\sim 0.2$) in order 
to ensure the deviation of 
$\theta_{\rm sol}$ from $\pi/4$.
At the same time
$(\lambda_{12} + \lambda_{13})$ has to be 
smaller than $\sim 0.2$ in order 
to satisfy the limit on $|U_{e3}|$. 
These two conditions imply 
(for $\lambda_{12,13} > 0$)
a hierarchy of the form  
$\lambda_{12} \gg \lambda_{13}$. 
The remaining parameter is 
$\lambda_{23} \ls (\lambda_{12} + 
\lambda_{13})/2 \sim \lambda_{12}/2$.  

  In Fig.\ \ref{fig:cpcsmla} 
we show scatter plots of the 
values of $\lambda_{ij}$, 
obtained by requiring 
that the corresponding values of 
neutrino mixing parameters
$\tan^2 \theta_{\rm sol}$,
$\sin^2 2 \theta_{\rm atm}$ and
$|U_{e3}|^2$  lie within their 1$\sigma$ 
allowed ranges, Eq.\ (\ref{eq:range}).
One can see from 
the left panels in Fig.\ \ref{fig:cpcsmla}
that $\lambda_{12} \simeq (0.21 - 0.26)$, 
$\lambda_{23} \ltap 0.10$ and
$\lambda_{13} \ltap 0.03$. Thus, one has
$\lambda_{23} \ltap \lambda_{12}/2$ and
$\lambda_{13} \ltap \lambda^2_{12}/2$.
In Fig.\ \ref{fig:cpcsmla} 
we also display the resulting correlations 
between the neutrino mixing parameters 
for which the 1$\sigma$ allowed ranges  
were used. From the panels 
in the right column 
one sees that $\ts \gs 0.42$ and 
$|U_{e3}|^2 \gs 0.017$. 
Most of the points tend to lie 
at rather large values of \ts{} and $|U_{e3}|^2$. 
Taking the 3$\sigma$ range in Eq.\ (\ref{eq:range}) 
leads to the lower limits of 
$\ts \gs 0.35$ and of $|U_{e3}|^2 \gs 0.003$. 

  We consider next two 
representative (and to a certain
degree typical) forms of $\lam{}$.
A viable possibility is the existence of
``hierarchical'' relations between
$\lambda_{12}$, $\lambda_{23}$ and
$\lambda_{13}$. One can have, for instance,
$\lambda_{23} \sim \lambda^2_{12}$ and
$\lambda_{13} \sim \lambda^3_{12}$, with
$\lambda_{12} \equiv \lambda$ and, e.g.,
$\lambda \sim (0.20 - 0.30)$.
This implies the following form of \lam{}: 
%%%%%%%%%%%%%%%%%%%%%%%%%%%%%%%%%%
\bea 
U_{\lambda} \sim 
\small
\left( \bad 
1 - \lambda^2/2 & \lambda  & \lambda^3 \\[0.2cm] 
- \lambda & 1 - \lambda^2/2 & \lambda^2 \\[0.2cm] 
\lambda^3 & - \lambda^2 & 1 
\ea   \right) + {\cal O}(\lambda^4)~.
\label{Ulamh}
\eea
%%%%%%%%%%%%%%%%%%%%%%%%%%%%%%%%%%%
%
\noindent The structure of \lam{} is  
close to that of a diagonal matrix, 
and is similar to the structure of the CKM matrix. 
Given the hierarchy in the 
charged lepton masses,
the CKM-like form of \lam{}, Eq.\ (\ref{Ulamh}), 
is rather natural.
One has in this case:
%%%%%%%%%%%%%%%%%%%%%%%%%%%%%%%%%%%%%%%%%%
\be \label{eq:mixcpcexp}
\ba 
\tan^2 \theta_{\rm sol}  \simeq 
1 - 2 \sqrt{2} \, \lambda  + 4 \, 
\lambda^2 - 2 \sqrt{2} \, \lambda^3~, \\[0.3cm]
|U_{e3}| \simeq \frac{\D \lambda}{\D \sqrt{2}}~, \\[0.3cm]
% includes the right sign
% U_{e3} \simeq - \frac{\D \lambda}{\D \sqrt{2}}~, \\[0.3cm]
\sin^2 2 \theta_{\rm atm} \simeq 1 - 4 \, \lambda^4  
\ea 
\ee
%%%%%%%%%%%%%%%%%%%%%%%%%%%%%%%%%%%%%%%%%%%
%
plus terms of higher order in $\lambda$. 
Consequently, this scenario 
``predicts'' solar neutrino mixing 
and $|U_{e3}|^2$ close to their  
currently allowed maximal values, 
and atmospheric neutrino mixing 
close to maximal.

   More specifically, the hierarchy 
of the $\lambda$--parameters we are considering,  
$\lambda_{12} \gg \lambda_{23, 13}$, 
leads, as it follows from 
Eqs.\ (\ref{eq:mixcpc}) and 
(\ref{eq:mixcpcexp}), 
to the interesting correlation:  
%%%%%%%%%%%%%%%%%%%%%%%%%%%%%%%%%%%%%%%%
\be \label{eq:corhie}
\tan^2 \theta_{\rm sol} \simeq 1 - 4 \, |U_{e3}| + 8 \, |U_{e3}|^2 - 
8 \, |U_{e3}|^3 ~. 
\ee
%%%%%%%%%%%%%%%%%%%%%%%%%%%%%%%%%%%%%%%%%
%
The  upper limit of 
$\tan^2 \theta_{\rm sol} \leq 0.52$ (0.72) 
implies in this case a lower limit 
on $|U_{e3}|^2 \gtap 0.027~(0.006)$~\footnote{That the upper limit on 
$\tan^2 \theta_{\rm sol}$  implies a significant lower limit 
on $|U_{e3}|^2$ in the case of 
CKM--like form of \lam{} was noticed 
also in \cite{GTani1}.}. 
This explains qualitatively 
the lower limit on $|U_{e3}|^2$ 
as observed in Fig.\ \ref{fig:cpcsmla} 
and mentioned earlier. 
Another correlation related to the CKM--like form 
of \lam{}, Eq.\ (\ref{Ulamh}), reads:  
%%%%%%%%%%%%%%%%%%%%%%%%%%%%%%%%%%%
\be
\sin^2 2 \theta_{\rm atm} \simeq 1 - 16 \, |U_{e3}|^4~.
\ee
%%%%%%%%%%%%%%%%%%%%%%%%%%%%%%%%%%
%
\noindent 
Thus, the deviations from maximal atmospheric 
neutrino mixing 
are determined in the case of the hierarchical
relations 
$\lambda_{12} \equiv \lambda \gg \lambda_{23,13}$,
$\lambda_{23} \sim \lambda^2$,
by the magnitude of $|U_{e3}|^4$.
For, e.g., $|U_{e3}|^2 = 0.029~(0.05)$, we have
$\sin^2 2 \theta_{\rm atm} \simeq 0.975~(0.96)$.  

  The alternative possibility is that
of a mild ``hierarchy'' between $\lambda_{12}$ and 
$\lambda_{23}$. We can have, for instance, 
$\lambda_{23} \simeq \lambda_{12}/2$,
$\lambda_{12} \equiv \lambda$.
Taking also
$\lambda_{13} \simeq \lambda^3 \ll \lambda_{12}$,
one finds:
%%%%%%%%%%%%%%%%%%%%%%%%%%%%%%%%%%%
\be \label{eq:mixcpcexp1}
\ba 
\tan^2 \theta_{\rm sol}  \simeq 
1 - 2 \sqrt{2} \, \lambda  + 
(4 - \sqrt{2}) \, \lambda^2  - 2 \sqrt{2} \, \lambda^3 ~, \\[0.3cm]
|U_{e3}| \simeq  \frac{\D \lambda}{\D \sqrt{2}} ~,\\[0.3cm]
% includes the right sign
% U_{e3} \simeq - \frac{\D \lambda}{\D \sqrt{2}} ~, ??? \\[0.3cm]
%
\sin^2 2 \theta_{\rm atm} \simeq 1 - \lambda^2 
\simeq 1 -  2\, |U_{e3}|^2 ~. 
\ea 
\ee
%%%%%%%%%%%%%%%%%%%%%%%%%%%%%%%%%%%%%%%%
%
The atmospheric neutrino mixing 
angle $\theta_{\rm atm}$ can 
deviate more from $\pi/4$ than in the 
hierarchical case:
for $|U_{e3}|^2 = 0.05$ we have now
$\sin^2 2 \theta_{\rm atm} \simeq 0.90$.  
This provides a possibility
to distinguish the case 
of $\lambda_{23} \simeq \lambda_{12}/2$,
$\lambda_{12,23} \gg \lambda_{13}$,
from the  hierarchical one 
$\lambda_{13} \sim \lambda^3_{12}
< \lambda_{23} \sim \lambda^2_{12} < \lambda_{12}$
considered above.
For  $\lambda \sim 0.24$ one finds 
from Eq.\ (\ref{eq:mixcpcexp1})
$\tan^2 \theta_{\rm sol}  \simeq 0.43$,
which is quite encouraging 
since values of $\lambda \sim 0.24$ 
could be interpreted in terms of 
$\sqrt{m_\mu/m_\tau} \simeq 
\sqrt{0.105/1.77} \simeq 0.24$. 

\vspace{-0.4cm}
%%%%%%%%%%%%%%%%%%%%%%%%%%%%%%%%%

\subsection{\label{sec:lam23}The Case of $\lambda_{23} = 1$}

%%%%%%%%%%%%%%%%%%%%%%%%%%%%%%%%%%
\vspace{-0.2cm}
\hskip 0.8cm If $\lambda_{23} = 1$ and $\lambda_{12,13}$ are ``small'',
$\lambda_{12,13} \ltap 0.35$, we find: 
%%%%%%%%%%%%%%%%%%%%%%%%%%%%%%%%%%
\be \label{eq:lam23=1}
\ba 
\tan^2 \theta_{\rm sol} \simeq 
1 - 2 \sqrt{2} \, (\lambda_{12} + \lambda_{13}) + 4 \, 
(\lambda_{12} + \lambda_{13})^2 + {\cal O}(\lambda^3) ~, \\[0.3cm]
U_{e3} \simeq \frac{\D \lambda_{12} - \lambda_{13}}{\D \sqrt{2}}  
+ {\cal O}(\lambda^3) ~, \\[0.3cm]
\sin^2 2 \theta_{\rm atm}  \simeq 1  - \frac{1}{4} \, 
(\lambda_{12}^2 + \lambda_{13}^2)^2 + \lambda_{12} \,  \lambda_{13} 
\, (\lambda_{12}^2 - \lambda_{13}^2) 
+ {\cal O}(\lambda^5)~.
\ea 
\ee
%%%%%%%%%%%%%%%%%%%%%%%%%%%%%%%%%%%%%
%
The correlations of 
$\lambda_{12}$ and $\lambda_{13}$  
as well as  of the mixing parameters 
for $\lambda_{23}=1$ are shown in 
Fig.\ \ref{fig:cpclam231}. 
We used the 1$\sigma$ allowed ranges of 
the oscillation parameters, 
Eq.\ (\ref{eq:range}).
As Fig.\ \ref{fig:cpclam231} shows,
$\sin^2 2 \theta_{\rm atm}$ is practically 1 in this case, 
$\sin^2 2 \theta_{\rm atm} \gs 0.9965$. 
The lower limit is reached for
relatively large $|U_{e3}|^2 \simeq 0.025$ and 
relatively small 
$\tan^2 \theta_{\rm sol} \simeq 0.32$.
However, such small deviations of 
$\sin^2 2 \theta_{\rm atm}$ from 1 
are extremely difficult to measure
\footnote{The requisite precision
may be achieved in experiments
at neutrino factories \cite{nufac}.}. 
For $\lambda_{23}=1$ and ``small'' 
$\lambda_{12, 13}$, \lam{} has the following
structure:
%%%%%%%%%%%%%%%%%%%%%%%%%%%%%%%%%%%%%%%%
\be 
\lam \simeq  \left( \bad 
1 - \frac{1}{2} \, (\lambda_{12}^2 + \lambda_{13}^2) & 
 \lambda_{12}  & \lambda_{13} \\[0.2cm] 
-\lambda_{13}   & - \lambda_{12} \, \lambda_{13} & 
1 - \frac{1}{2} \,  \lambda_{13}^2  \\[0.2cm] 
\lambda_{12}   & -1 + \frac{1}{2} \, \lambda_{12}^2   &  0 
\ea   \right) + {\cal O}(\lambda^3)~.
\label{Ulam1}
\ee
%%%%%%%%%%%%%%%%%%%%%%%%%%%%%%%%
%
All three possibilities $\lambda_{12} \gg \lambda_{13}$, 
$\lambda_{13} \gg \lambda_{12}$ and $\lambda_{12} \sim \lambda_{13}$
are allowed. In any of these three cases
the 23 submatrix of \lam{} displays an 
antidiagonal structure, as seen in Eq.\ (\ref{Ulam1}).

%%%%%%%%%%%%%%%%%%%%%%%%%%%%%%%%%%%%%%%%%%%%%%%%%%%

\subsection{\label{sec:large}Large $\lambda_{ij}$}

%%%%%%%%%%%%%%%%%%%%%%%%%%%%%%%%%%%%%%%%%%%%%%%%%%%%
%
\hskip 0.8cm It is also possible that all 
$\lambda_{ij} = \sin \theta'_{ij}$ 
are relatively large,
$\lambda_{ij} \gtap 0.40$. 
From Fig.\ \ref{fig:mostgencpc}  
one sees that in this case 
$\lambda_{12} \simeq (0.92 - 0.99)~[0.85-0.99]$, 
$\lambda_{13} \simeq (0.57 - 0.77)~[0.47-0.85]$, 
and $\lambda_{23} \simeq (0.57 - 0.92)~[0.40-0.99]$, 
when the 1 s.d.\ [3 s.d.] ranges of the oscillation parameters 
are used. It turns out that, $\cos \theta'_{23} > 0$, whereas 
$\cos \theta'_{12, 13}$ can take both signs. 

    We limit our discussion to the 
case of all $\lambda_{ij}$ being ``large'': 
$\lambda_{ij} \geq 1/\sqrt{2}$.
We can use the deviations of 
$\lambda_{12}$ from 1, 
and the deviations
of $\lambda_{13,23}$ from $1/\sqrt{2}$, 
as small parameters
to get convenient expressions 
for $\tan^2\theta_{\rm sol}$, 
$|U_{e3}|^2$ and $\sin^2 2 \theta_{\rm atm}$:
%%%%%%%%%%%%%%%%%%%%%%%%%%%%%%%%%%%%%%%
\bea
\lambda_{12} \equiv 1 - \epsilon_{12}^2 \\[0.3cm]
\lambda_{13} \equiv \frac{1}{\sqrt{2}} \, (1 + \epsilon_{13}^2) \\[0.3cm]
\lambda_{23} \equiv \frac{1}{\sqrt{2}} \, (1 + \epsilon_{23}^2)~,
\label{eq:eij}
\eea
%%%%%%%%%%%%%%%%%%%%%%%%%%%%%%%%%%%%%%%%%%%%
%
where $\epsilon^2_{12} \ltap 0.08~(0.15)$,
$\epsilon^2_{23} \ltap 0.30~(0.40)$ and
$\epsilon^2_{13} \ltap 0.09~(0.20)$, where the 1 s.d.\ (3 s.d.) 
ranges of the oscillation parameters were used.  
One obtains for
sufficiently small $\epsilon_{ij}$:
%%%%%%%%%%%%%%%%%%%%%%%%%%%%%%%%%%%%%%%%%%%%%%%
\be 
\ba
\tan^2 \theta_{\rm sol} \simeq 
 1 - 4 \, \epsilon_{12} + 8 \, \epsilon_{12}^2 + {\cal O}(\epsilon^3)~, 
\\[0.3cm] 
U_{e3} \simeq \epsilon_{23}^2 -  \epsilon_{12} + {\cal O}(\epsilon^3)~, 
\\[0.3cm] 
\sin^2 2 \theta_{\rm atm}  \simeq 1 + {\cal O}(\epsilon^4)~.
\label{hier}
\ea 
\ee
%%%%%%%%%%%%%%%%%%%%%%%%%%%%%%%%%%
%
The terms of higher order in $\epsilon$ 
we do not give can be sizable.
The scatter plots do not show any 
new important correlation. 
In particular,
$\sa{} \simeq 0.95$ is possible 
also in the case under discussion.
Both $\epsilon_{23}$ and 
$\epsilon_{13}$ can be zero, 
whereas $\epsilon_{12}$ cannot. 
No hierarchy is implied. 
If $\epsilon_{12} \gg \epsilon_{13,23}$, we get
$|U_{e3}| \simeq \epsilon_{12}(1 - \epsilon^2_{12}/4) + {\cal O}(\epsilon^4)$,
and
%%%%%%%%%%%%%%%%%%%%%%%%%%%%%%%%%%%%%%%%%%%%%%%%%
\be 
\tan^2 \theta_{\rm sol} \simeq 
 1 - 4~|U_{e3}|  + 8~|U_{e3}|^2 - 16~|U_{e3}|^3~.
\label{eps12}
\ee
%%%%%%%%%%%%%%%%%%%%%%%%%%%%%%%%%%%%%%%%%%%%%%%%
%

  For all $\lambda_{ij}$ ``large'',
$\lambda_{ij} \geq 1/\sqrt{2}$,
\lam{} has the form:
%%%%%%%%%%%%%%%%%%%%%%%%%%%%%%%%%%%%%%%%
\be  \label{eq:lamla}
\lam \simeq  \left( \bad 
\epsilon_{12} & 
\frac{1}{\sqrt{2}} - \frac{\epsilon_{12}^2 + \epsilon_{13}^2}{\sqrt{2}}
& \frac{1}{\sqrt{2}} + \frac{\epsilon_{13}^2}{\sqrt{2}}\\[0.2cm] 
\frac{-1}{\sqrt{2}} - \frac{\epsilon_{12}}{\sqrt{2}} + 
\frac{\epsilon_{12}^2 + \epsilon_{23}^2}{\sqrt{2}} & 
\frac{-1}{2} + \epsilon_{12} + \frac{1}{2} 
\, (\epsilon_{12}^2 - \epsilon_{13}^2 - \epsilon_{23}^2) & 
\frac{1}{2} + \frac{1}{2} \, (\epsilon_{23}^2 - \epsilon_{13}^2)\\[0.2cm] 
\frac{1}{\sqrt{2}} - \frac{\epsilon_{12}}{\sqrt{2}} + 
\frac{\epsilon_{23}^2 - \epsilon_{12}^2}{\sqrt{2}} & 
\frac{-1}{2} - \epsilon_{12} + \frac{1}{2} 
\, (\epsilon_{12}^2 - \epsilon_{13}^2 + \epsilon_{23}^2) & 
\frac{1}{2} - \frac{1}{2} \, (\epsilon_{23}^2 + \epsilon_{13}^2)
\ea   \right) + {\cal O}(\epsilon^3)~.
\ee
%%%%%%%%%%%%%%%%%%%%%%%%%%%%%%%%%%%%%%%%%%%%%%
%
All entries in \lam{}, except the 11 entry, 
are large.\\ 

 Let us summarize the results obtained 
under the assumption of CP--conservation: 
\begin{itemize}
\item If all $\lambda_{ij}$ are small, one 
typically has $\ts \gs 0.42 $ and $|U_{e3}|^2 \gs 0.02$.   
The matrix \lam{} has a ``CKM--like'' structure. 
One can have $\sa \simeq 0.95$, i.e., \sa{} can
deviate noticeably from 1.

\item If $\lambda_{23} = 1$ and $\lambda_{12,13}$ are small, 
one practically has $\sa{} \simeq 1$, the possible deviations
from 1 being exceedingly small.
Any deviation of $\sa{}$ from 1
would imply rather large $|U_{e3}|^2$ and 
relatively small \ts. The 23 block of the matrix \lam{} 
has an antidiagonal (quasi--Dirac like) form. 

\item If all $\lambda_{ij}$ are large,
$\lambda_{ij} \gtap 1/\sqrt{2}$,  
the matrix \lam{} displays the 
unusual structure given in Eq.\ (\ref{eq:lamla}),
with all elements, except the 11 element,
being large. Also in this case we can  
have $\sa{} \simeq 0.95$.

\end{itemize}

\vspace{-0.4cm}
%%%%%%%%%%%%%%%%%%%%%%%%%%%%%%%%%%%%

\subsection{Implied Forms of the Charged Lepton Mass Matrix}

%%%%%%%%%%%%%%%%%%%%%%%%%%%%%%%%%%%%%
\vspace{-0.2cm}
 
\hskip 0.8cm The matrix \lam{} arises from the 
diagonalization of the 
charged lepton mass matrix $m_{\rm lep}$. 
Having the approximate form of \lam{}
and using the knowledge 
of the charged lepton masses $m_e$, $m_\mu$ and 
$m_\tau$, we can derive the structure of 
two different matrices related to $m_{\rm lep}$. 
The first one is the matrix $m_{\rm lep}~m_{\rm lep}^\dagger$, which is 
diagonalized by \lam, i.e., 
%%%%%%%%%%%%%%%%%%%%%%%%%%%%%%%%
\be 
\label{eq:ml1}
m_{\rm lep}~m_{\rm lep}^\dagger = \lam~(m_{\rm lep}^{\rm diag})^2~\lam^\dagger~,
\ee 
%%%%%%%%%%%%%%%%%%%%%%%%%%%%
%
where $m_{\rm lep}^{\rm diag}$ is a diagonal mass matrix containing the 
charged lepton masses $m_e$, $m_\mu$ and $m_\tau$. 
The second matrix is 
%%%%%%%%%%%%%%%%%%%%%%%%%%%%%
\be
\label{eq:ml2}
\lam~m_{\rm lep}^{\rm diag}~\lam^T~.
\ee
%%%%%%%%%%%%%%%%%%%%%%%%%%%%%%%%%
%
Under the assumption of $m_{\rm lep}$ being symmetric,
which is realized in some GUT theories,
the matrix in Eq.\ (\ref{eq:ml2}) 
coincides with the charged lepton mass matrix. 
 
  Consider first the matrix $m_{\rm lep}~m_{\rm lep}^\dagger$. 
In the case of ``small'' 
$\lambda_{ij}$, we find that
%%%%%%%%%%%%%%%%%%%%%%%%%%%%%%%%%%%%%%%%%
\be 
m_{\rm lep}~m_{\rm lep}^\dagger \simeq 
\left( 
\bad 
m_e^2 + m_\tau^2~\lambda_{13}^2 + m_\mu^2~\lambda_{12}^2 & 
m_\tau^2~\lambda_{13}~\lambda_{23} + m_\mu^2~\lambda_{12} & 
m_\tau^2~\lambda_{13} - m_\mu^2~\lambda_{12}~\lambda_{13} \\[.3cm]
\cdot & 
m_\mu^2 + m_\tau^2~\lambda_{23}^2 & 
m_\tau^2 \lambda_{23} - m_\mu^2~\lambda_{12}~\lambda_{13}\\[.3cm]
\cdot & \cdot & m_\tau^2
\ea
\right)~
\ee
%%%%%%%%%%%%%%%%%%%%%%%%%%%%%%%%%%%%%%%%%%%
%
plus terms of order $\lambda^3$. 
Thus, $m_{\rm lep}~m_{\rm lep}^\dagger$ is close
to a diagonal matrix.  
In the special case 
of $\lambda_{13} = \lambda_{23} \simeq 0$ we get: 
%%%%%%%%%%%%%%%%%%%%%%%%%%%%%%%%%
\be 
m_{\rm lep}~m_{\rm lep}^\dagger \simeq 
\left( 
\bad 
m_e^2 + m_\mu^2~\lambda^2 & 
m_\mu^2~\lambda & 0 \\[.3cm]
m_\mu^2~\lambda & 
m_\mu^2 & 0 \\[.3cm]
0 & 0 & m_\tau^2
\ea
\right)~.
\ee
%%%%%%%%%%%%%%%%%%%%%%%%%%%%%%%%%%%
%
The zero entries can be terms of order $m^2_e$. The matrix
$m_{\rm lep}~m_{\rm lep}^\dagger$ has clearly a hierarchical structure.\\

  When $\lambda_{23}=1$ and 
$\lambda_{12,13}$ are  ``small'', one obtains:
%%%%%%%%%%%%%%%%%%%%%%%%%%%%%%%
\be 
m_{\rm lep}~m_{\rm lep}^\dagger \simeq 
\left( 
\bad 
m_e^2 + m_\tau^2~\lambda_{13}^2 + m_\mu^2~\lambda_{12}^2 & 
m_\tau^2~\lambda_{13}  & 
- m_\mu^2~\lambda_{12} \\[.3cm]
m_\tau^2~\lambda_{13} & 
m_\tau^2 & 
m_\mu^2~\lambda_{12}~\lambda_{13}\\[.3cm]
- m_\mu^2~\lambda_{12} & m_\mu^2~\lambda_{12}~\lambda_{13}  &m_\mu^2
\ea
\right)~.
\ee
%%%%%%%%%%%%%%%%%%%%%%%%%%%%%%%%%%%%
%
In contrast to the previous case, now
the 22 element is equal to
$m_\tau^2$ and the 33 element
is given by $m_\mu^2$.
The 12 block of 
$m_{\rm lep}~m_{\rm lep}^\dagger$ is hierarchical,
while the 23 block has ``inverted hierarchy--like'' form.

 Finally, if all $\lambda_{ij}$ are ``large'', 
we can expect that $m_{\rm lep}~m_{\rm lep}^\dagger$ 
is to a good approximation a ``democratic'' matrix. 
Indeed, one finds:
%%%%%%%%%%%%%%%%%%%%%%%%%%%%%%%%%%%%%%%%
\be 
m_{\rm lep}~m_{\rm lep}^\dagger \simeq \frac{m_\tau^2}{2}~
\left( 
\bad 
1 & 1/\sqrt{2} & 1/\sqrt{2} \\[0.3cm]
1/\sqrt{2} & 1/2 & 1/2  \\[0.3cm]
1/\sqrt{2} & 1/2 & 1/2
\ea
\right)
\ee
%%%%%%%%%%%%%%%%%%%%%%%%%%%%%
%
plus terms of order $\epsilon$.\\

   From the definition of the two characteristic 
mass matrices in Eqs.\ (\ref{eq:ml1}) and (\ref{eq:ml2}) 
it follows that the expressions for $\lam~m_{\rm lep}^{\rm diag}~\lam^T$ 
in the three cases we are discussing 
can be obtained from the corresponding 
expressions of $m_{\rm lep}~m_{\rm lep}^\dagger $ given above 
by replacing $m_l^2$ with $m_l$ in the latter. 

\vspace{-0.4cm}
%%%%%%%%%%%%%%%%%%%%%%%%%%%%%%%%%%%

\section{\label{sec:CPV}CP Nonconservation}

%%%%%%%%%%%%%%%%%%%%%%%%%%%%%%%%%%%%%%%%%%%%%%%%%
%%%%%%%%%%%%%%%%%%%%%%%%%%%%%%%%%%%%%%%%%%%%%%%%%%%%%%%%

\subsection{\label{sec:cpvgen}General Considerations}

%%%%%%%%%%%%%%%%%%%%%%%%%%%%%%%%%%%%%%%%
\vspace{-0.2cm}
\hskip 0.8cm It is well known 
that if the neutrinos with definite 
mass are Majorana particles,
the PMNS matrix contains 6 physical 
parameters, three mixing angles, 
and three CP--violating phases. 
The latter can be 
divided in one ``Dirac phase'', 
$\delta$, and two ``Majorana phases'', 
$\alpha$ and $\beta$. The Majorana phases 
appear only in amplitudes 
describing lepton number 
violating processes in which the 
total lepton charge changes by two units.
In general, as we have already discussed, one has

%%%%%%%%%%%%%%%%%%%%%%%%%%%%%%%%%%%%%%%%%%
\be
\pmns = U_{\rm lep}^\dagger \, U_\nu~.
\label{UPMNSCPV}
\ee
%%%%%%%%%%%%%%%%%%%%%%%%%%%%%%%%%%%%%%%%%%%
%
where $U_{\rm lep}$ and $U_\nu$ two $3\times 3$ 
unitary matrices: $U_{\rm lep}$ arises from the 
diagonalization of the 
charged lepton mass matrix, 
while $U_\nu$ diagonalizes the 
neutrino Majorana mass term.
Any $3\times 3$  unitary matrix contains 
3 moduli and 6 phases and 
can be written as \cite{wir}
%%%%%%%%%%%%%%%%%%%%%%%%%%%%%%%%%%%%%%%
\be \label{eq:unit}
U = e^{i \Phi} \, P \, \tilde{U} \, Q~, 
\ee
%%%%%%%%%%%%%%%%%%%%%%%%%%%%%%%%%%%%%%%%%%%%%
%
where $P \equiv {\rm diag} (1,e^{i \phi_1},e^{i \phi_2})$
and $Q \equiv {\rm diag} (1,e^{i \rho_1},e^{i \rho_2}) $ 
are diagonal phase matrices having 
2 phases each, and $\tilde{U}$ is a unitary 
``CKM--like'' matrix containing 1 phase and 3 angles. 
The charged lepton Dirac mass term, 
$m_{\rm lep}$, 
is diagonalized by a bi--unitary transformation:
%%%%%%%%%%%%%%%%%%%%%%%%%%%%%%%%%%%%%%%%%%%%%
\be
m_{\rm lep} = U_L \, m_{\rm lep}^{\rm diag} \, U_R^\dagger ~,
\ee
%%%%%%%%%%%%%%%%%%%%%%%%%%%%%%%%%%%%%%%%%%%%
%
where $U_{L,R}$ are $3\times 3$ unitary matrices
and $m_{\rm lep}^{\rm diag}$ is the diagonal matrix
containing the masses of the charged leptons.
Casting $U_{L,R}$ in the form (\ref{eq:unit}), i.e., 
$U_{L,R} = e^{i \Phi_{L,R}} \, P_{L,R} \, \tilde{U}_{L,R} \, Q_{L,R}$, 
we find 
%%%%%%%%%%%%%%%%%%%%%%%%%%%%%%%%%%%%%%%%%%%%
\be
m_{\rm lep} = e^{i (\Phi_L - \Phi_R)} \, Q_L \, 
\tilde{U}_L \, P_L \, m_{\rm lep}^{\rm diag} \, 
Q_R^\dagger \, \tilde{U}_R^\dagger \, P_R^\dagger~.
\ee
%%%%%%%%%%%%%%%%%%%%%%%%%%%%%%%%%%%%%%%%%%%%
%
The term $P_L \, m_{\rm lep}^{\rm diag} \, Q_R^\dagger$ contains 
only 2 relative phases, which can be 
associated with the right--handed charged
lepton fields.
The three independent phases in   
$e^{i (\Phi_L - \Phi_R)} \, Q_L$ can be 
absorbed by a redefinition of the 
left--handed charged lepton fields. 
Therefore, $U_{\rm lep}$ is effectively given  
by $\tilde{U}_L$ 
and contains three angles and one phase. 

   The neutrino mass matrix $m_\nu$ is diagonalized via 
%%%%%%%%%%%%%%%%%%%%%%%%%%%%%%%%%%%%%%%%%
\be
m_\nu = U_\nu \, m_\nu^{\rm diag} \, U_\nu^T ~.
\ee
%%%%%%%%%%%%%%%%%%%%%%%%%%%%%%%%%%%%%%%%
%
The unitary matrix $U_\nu$ can be written  
in the form (\ref{eq:unit}).
It is not possible to 
absorb phases in the neutrino 
fields since the neutrino mass term is 
of Majorana type \cite{BHP80,Doi81}. Thus,  
%%%%%%%%%%%%%%%%%%%%%%%%%%%%%%%%%%%%%%%%%%%%%%
\be \label{eq:pmnscpv}
\pmns = U_{\rm lep}^\dagger \, U_\nu = e^{i \Phi \nu} \, 
\tilde{U}_{\rm lep}^\dagger \, P_\nu \, \tilde{U}_\nu \, Q_\nu ~.
\ee
%%%%%%%%%%%%%%%%%%%%%%%%%%%%%%%%%%%%%
%
The common phase $\Phi_\nu$ has no physical meaning and  
we will ignore it. 
Consequently, in the most general case,  
the elements of
\pmns{} given by Eq.\ (\ref{eq:pmnscpv}) 
are expressed in terms of six real 
parameters and six phases
in $\tilde{U}_{\rm lep}$ and $U_\nu$~
\footnote{Using a different representation 
of the unitary matrices
$U_{\rm lep}$ and $U_\nu$, the authors of \cite{GTani2} 
came to the conclusion that \pmns{} 
is expressed in terms of 
six real parameters and seven phases
of $U_{\rm lep}$ and $U_\nu$. 
Writing the unitary matrices in the form of
Eq.\ (\ref{eq:unit}) 
allows to reduce the number of phases by one.}. 
Only six combinations of those ---
the three angles and the three phases of
\pmns{}, are observable,
in principle, at low energies.
Note that the two phases 
in $Q_\nu$ are ``Majorana--like'', 
since they will not appear in 
the probabilities describing 
the flavour neutrino oscillations 
\cite{BHP80,Lang86}. Note also that 
if $U_{\rm lep} = {\mathbbm 1}$,
the phases in the matrix $P_\nu$ can be
eliminated by a redefinition of the     
charged lepton fields. 

  The requirement of bimaximality 
of $\tilde{U}_\nu$ implies 
that $\tilde{U}_\nu$ is real 
and given by Eq.\ (\ref{eq:Ubimax}). 
In this case the three angles 
and the Dirac phase in the PMNS matrix \pmns{}
will depend 
in a complicated manner
on the three angles and the phase in 
$\tilde{U}_{\rm lep}$ and on 
the two phases in $P_\nu$. The two Majorana phases will depend 
in addition on the parameters in $Q_\nu$. 

  It should be emphasized that the form of \pmns{} 
given in Eq.\ (\ref{eq:pmnscpv})
is the most general one. 
A specific model
in the framework of which the bimaximal 
$\tilde{U}_\nu$ is obtained, might 
imply symmetries or textures, along the lines of, 
e.g., \cite{GFMarf}, in 
$m_\nu$, which will reduce the number 
of independent parameters in $U_\nu$.
We will comment in Section \ref{sec:spec} on 
such possibilities. 

  In the scheme with three massive 
Majorana neutrinos under discussion there exist 
three rephasing invariants related
to the three CP--violating phases
in ${\rm U_{\rm PMNS}}$, $\delta$,
$\alpha$ and $\beta$ \cite{CJ85,PKSP3nu88,JMaj87,BrancoLR86,ASBranco00}. 
The first
is the standard Dirac one $J_{CP}$ \cite{CJ85}, associated
with the Dirac phase $\delta$:
%%%%%%%%%%%%%%%%%%%%%%%%%%%%%%%%%%%%%%%%%%%%%%%%%%%
\be 
\label{eq:JCP}
J_{CP} = 
{\rm Im} \left\{ U_{e1} \, U_{\mu 2} \, U_{e 2}^\ast \, U_{\mu 1}^\ast 
\right\}~. 
\ee
%%%%%%%%%%%%%%%%%%%%%%%%%%%%%%%%%%%%%%%%%%%%%%%%%%%
%
It determines the magnitude 
of CP--violation effects in neutrino oscillations
\cite{PKSP3nu88}. Let us note that if $U_{\rm lep} = {\mathbbm 1}$
and $\tilde{U_\nu}$ is a real matrix, one has $J_{CP} = 0$. 

   The two additional
invariants, $S_1$ and $S_2$, whose existence is
related to the Majorana nature of massive 
neutrinos, i.e., to the phases 
$\alpha$ and $\beta$, can be chosen as \cite{JMaj87,ASBranco00} 
(see also \cite{BPP1})
\footnote{We assume that the fields
of massive Majorana neutrinos satisfy
Majorana conditions which 
do not contain phase factors.}:
%%%%%%%%%%%%%%%%%%%%%%%%%%%%%%%%%%%%
\be
S_1 = {\rm Im}\left\{ U_{e1} \, U_{e3}^\ast \right\}~,~~~~
S_2 = {\rm Im}\left\{ U_{e2} \, U_{e3}^\ast \right\}~.
\label{eq:SCP}
\ee
%%%%%%%%%%%%%%%%%%%%%%%%%%%%%%%%%%%%%%%
%
If $S_1 \neq 0$ and/or $S_2 \neq 0$, CP is not conserved
due to the Majorana phases $\beta$ and/or $(\alpha - \beta)$.
The effective Majorana mass in 
$\betabeta$--decay, $\meff$, depends, in general, on 
$S_1$ and $S_2$ \cite{BPP1} and not on $J_{CP}$. 
Let us note, however, even if $S_{1,2} = 0$ 
(which can take place if, e.g., $|U_{e3}| = 0$),
the two Majorana phases $\alpha$ and $\beta$
can still be a source of CP--nonconservation
in the lepton sector provided 
${\rm Im}\left\{ U_{e1} \, U_{e2}^\ast \right\} \neq 0$
and ${\rm Im}\left\{ U_{\mu 2} \, U_{\mu 3}^\ast \right\}\neq 0$
\cite{ASBranco00,JNieves02}. 

  Let us denote the phase in 
$\tilde{U}_{\rm lep}$ by $\psi$. 
We will include it 
in $\tilde{U}_{\rm lep}$  
in the same way 
this is done for
the phase $\delta$ 
in Eq.\ (\ref{eq:Upara}). 
We will also 
use $P_\nu = {\rm diag} (1,e^{i \phi},e^{i \omega})$
and $Q_\nu \equiv {\rm diag} (1,e^{i \rho},e^{i \sigma})$. 
This means that the Dirac 
phase $\delta$, which has observable 
consequences in neutrino oscillation 
experiments, 
is determined {\it only by the phases 
$\psi$, $\phi$  and $\omega$}. 
The Majorana phases in \pmns{},
$\alpha$ and $\beta$, 
receive contributions also from the 
two remaining phases $\rho$ and $\sigma$.  
Allowing the phases $\delta$, $\alpha$ and $\beta$
to vary between 0 and $2 \pi$, permits
to constrain (without loss of generality)
the mixing angles 
$\theta_{ij}$ to lie between 0 and $\pi/2$. 
In Fig.\ \ref{fig:cpv} we show the result of 
a random search for the allowed regions
of the values of the parameters $\lambda_{ij}$. 
The regions
present in Fig.\ \ref{fig:mostgencpc}
are again allowed. Clustering of points 
around these regions 
is also observed. The effect of the phases is that 
values of the parameters, located
in areas between the regions corresponding to 
the case of CP--conservation are allowed. 
Let us consider the physical observables of interest
in the three main regions of the parameter space we
have identified earlier. 

%%%%%%%%%%%%%%%%%%%%%%%%%%%%%%%%%%%%

\subsection{\label{sec:smla}Small $\lambda_{ij}$}

%%%%%%%%%%%%%%%%%%%%%%%%%%%%%%%%%%%%%

\hskip 0.8cm Let us choose 
$\lambda_{12} = \lambda$, $\lambda_{23} = A \, \lambda$ and 
$\lambda_{13} = B \, \lambda$ with $A, B$ 
real and of order one
\footnote{Hierarchical $\lambda_{ij}$ correspond in this 
parametrization to sufficiently small $A$ and/or $B$.}. 
In this case $\tan^2 \theta_{\rm sol}$, $U_{e3}$ and
$\sin^2 2 \theta_{\rm atm}$ are given by 
%%%%%%%%%%%%%%%%%%%%%%%%%%%%
\be \label{eq:mixcp}
\ba 
\tan^2 \theta_{\rm sol} \simeq 1 - 2 \sqrt{2} \, 
(c_\phi - B \, c_{\omega - \psi} ) \, \lambda +  
2 \left(2 \, (c_\phi - B \, c_{\omega - \psi})^2 - \sqrt{2} \, A 
\, (c_\omega + B \, c_{\phi - \psi}) \right) \, \lambda^2 ~,\\[0.3cm]
U_{e3} \simeq 
\sqrt{\frac{\D 1}{\D 2} (\D1+ B^2 + 2 \, B \, c_{\omega - \phi - \psi})} 
\left( \lambda~ - ~ 
\frac{\D (1 - B^2) \, A \, c_{\omega - \phi}}
{\D 1 + B^2 + 2 \, B \, c_{\omega - \phi - \psi}} \, \lambda^2 \right)~, 
\\[0.3cm]
\sin^2 2 \theta_{\rm atm}  \simeq  1 - 4 \, A^2  \, c_{\omega - \phi} \, 
\lambda^2 
\ea 
\ee
%%%%%%%%%%%%%%%%%%%%%%%%%%%%%%
%
plus terms of order ${\cal{O}}(\lambda^3)$. We introduced the 
obvious notation $c_\phi = \cos \phi$,  
$s_{\omega - \psi} = \sin (\omega - \psi)$, etc. 
Note that the possibility of all three $\lambda_{ij}$ being of the same 
order was not possible in case of conservation of CP.  
The rephasing invariant $J_{CP}$, 
which controls the magnitude of 
CP--violating effects in neutrino 
oscillations, has the form: 
%%%%%%%%%%%%%%%%%%%%%%%%%%%%%%%%%%%%%%%%
\be 
\label{eq:JCPsmla}
J_{CP} 
\simeq \frac{\lambda}{4 \sqrt{2}} \, 
\left ( s_\phi + B \, s_{\omega - \psi} + 
\lambda \, (s_{\omega - 2\phi} + B \, s_{2\omega - \phi - \psi}) \, A\right ) 
+ {\cal{O}}(\lambda^3)~.
\ee
%%%%%%%%%%%%%%%%%%%%%%%%%%%%%%%%%%%%%
%
Thus, {\it not only the phase $\psi$ of $\tilde{U}_{\rm lep}$,
but also the phases $\omega$ and $\phi$ from $P_\nu$
contribute to $J_{CP}$}. 
Using Eqs.\ (\ref{eq:mixcp}) and
(\ref{eq:JCPsmla}) it is not difficult to convince
oneself that the magnitude of $J_{CP}$ 
is controlled by the magnitude of $|U_{e3}|$. Indeed,
if, e.g., $c_{\omega - \phi - \psi} = -1$ and
$B=1$, so that the term $\sim \lambda$ in the expression
for  $U_{e3}$ vanishes and to leading order
$U_{e3} \simeq -\sqrt{2}\, A \, c_{\omega - \phi} \, \lambda^2$,
the term $\sim \lambda$ in $J_{CP}$ also vanishes and 
we have $J_{CP} \sim U_{e3}$. 

  The invariants $S_1$ and $S_2$ which 
(for $|U_{e3}| \neq 0$)
determine the magnitude of the 
effects of CP--violation 
associated with the Majorana 
nature of massive neutrinos, read:
%%%%%%%%%%%%%%%%%%%%%%%%%%%%%%%%%%%%
\bea
S_1 
% = {\rm Im}\left\{ U_{e1} \, U_{e3}^\ast \right\} 
\simeq \frac{\D \lambda}{\D 2} 
\left( s_{\phi + \sigma} + B \, s_{\omega - \psi + \sigma} 
\right) 
\\[0.3cm]
S_2
% = {\rm Im}\left\{ U_{e2} \, U_{e3}^\ast \right\} 
\simeq \frac{\D \lambda}{\D 2} 
\left( s_{\phi - \rho + \sigma} + B \, s_{\omega - \psi - \rho + \sigma} 
\right) ~,
\label{eq:S12small}
\eea
%%%%%%%%%%%%%%%%%%%%%%%%%%%%%%%%%%%%%
%
where we gave only the terms of order $\lambda$. 
{\it We see that all five phases in $\tilde{U}_{\rm lep}$, 
$P_\nu$ and $Q_{\nu}$, and not 
only the phases $\rho$ and $\sigma$ in $Q_{\nu}$,
contribute to $S_1$ and $S_2$}. In the leading order
expressions for $S_{1,2}$, Eq.\ (\ref{eq:S12small}),
the five phases enter in three 
independent combinations. Moreover, in the case of 
$\rho = \sigma = 0$, which is realized 
if the bimaximal mixing structure of $U_{\nu}$ 
is associated (in the limit of $U_{\rm lep} = {\mathbbm 1}$) 
with the approximate conservation of $L' = L_e - L_\mu - L_\tau$ 
(see Subsection \ref{sec:spec}), we have to 
leading order in $\lambda$: 
%%%%%%%%%%%%%%%%%%%%%%%%%%%%%%%%%%%%%%%%%%%
\be
J_{CP} \simeq \frac{S_1 }{2 \, \sqrt{2}} \simeq \frac{S_2 }{2 \, \sqrt{2}}~.
\label{eq:JS12}
\ee
%%%%%%%%%%%%%%%%%%%%%%%%%%%%%%%%%%%%%%%%%
%
Thus, {\it the magnitude of the CP--violating effects 
in neutrino oscillations is directly related in this case 
to the magnitude of the CP--violating effects associated
with the Majorana nature of neutrinos}. To leading order,
both types of CP--violating effects are due to two phases,
$\phi$ and $(\omega - \psi)$. 
We will consider a model in which 
$\rho = \sigma = 0$ in Subsection \ref{sec:spec}.

  In  Fig.\ \ref{fig:hie} we show 
the correlations between 
the observables
\ts{}, $|U_{e3}|^2$ and \sa,
when a hierarchy between the three
$\lambda$ parameters of the 
form we have considered
in Subsection \ref{sec:lam23},
$\lambda_{12} \equiv \lambda$, 
$\lambda_{13} = \lambda^3$ and 
$\lambda_{23} = \lambda^2$, is assumed.
The $1\sigma$ allowed  ranges of 
\ts{}, $|U_{e3}|^2$ and $\sin^2 2 \theta_{\rm atm}$,
were used to constrain $\lambda$
and the CP--violating phases.
As it follows from Fig.\ \ref{fig:hie}, 
\ts{} can take values 
in the whole $1\sigma$ 
interval allowed by the data, Eq.\ (\ref{eq:range}).
This is in sharp contrast 
to the case of CP--conservation,
in which the correlation between
the values $|U_{e3}|$ and 
the deviation of \ts{} from 1
leads, as Fig.\ \ref{fig:hie} shows, 
to the lower limit $\ts \gtap 0.48$.
We find also that if CP is not conserved, 
the following lower bound holds:
$|U_{e3}|^2 \gtap 0.018$.
There are interesting correlations
between \ts{}, $|U_{e3}|^2$ and \sa.
For instance, $\ts{} \leq 0.40$ 
implies $|U_{e3}|^2 \gs 0.026$ and  
$ \sa \ls 0.975$.

%%%%%%%%%%%%%%%%%%%%%%%%%%%%%%%%%%%%%%%%%%%

\subsection{\label{sec:lam23cpv} The Case of $\lambda_{23} = 1$}

%%%%%%%%%%%%%%%%%%%%%%%%%%%%%%%%%%%%%%%%%%%%%

\hskip 0.8cm This case corresponds to 
``small'' $\lambda_{12, 13}$, 
$\lambda_{12, 13} \ls 0.35$.  
Introducing $\lambda_{12} = \lambda$ and
$\lambda_{13} = B\lambda$, with $B$ real, we find:
%%%%%%%%%%%%%%%%%%%%%%%%%%%%%%%%%%%%%%%%%%%%%%
\be 
\ba 
\tan^2 \theta_{\rm sol} \simeq 1 - 2 \sqrt{2} \, 
(c_\omega + B \, c_{\phi - \psi} ) \, \lambda +  
4 \left(c_\omega + B \, c_{\omega - \psi} \right)^2 \, \lambda^2 ~,\\[0.3cm]
U_{e3} \simeq 
\lambda \, \sqrt{\frac{\D 1}{\D 2} (1 + B^2 - 2 \, B \, 
c_{\omega - \phi + \psi})} + {\cal O}(\lambda^3)~, \\[0.3cm]
\sin^2 2 \theta_{\rm atm}  \simeq  1 - 
\frac{1}{4} \, (1 - B^2 - 2 \, B \, c_{\omega - \phi - \psi})^2 \, 
\lambda^4 ~.
\ea 
\ee
%%%%%%%%%%%%%%%%%%%%%%%%%%%%%%%%%%%%%%%%%%%
%
The formulae for the mixing parameters
can actually be obtained
from the corresponding formulae
for all $\lambda_{ij}$ small, 
Eq.\ (\ref{eq:mixcp}), by setting $A = 0$ and 
making the change $B \rightarrow -B$, 
$\phi \leftrightarrow \omega$. 
The rephasing invariant $J_{CP}$ reads: 
%%%%%%%%%%%%%%%%%%%%%%%%%%%%%%%%%%%%%%%%%%%%%%%%%
\be
J_{CP} \simeq \frac{\lambda}{4 \sqrt{2}} \, \left( 
s_\omega - B \, s_{\phi - \psi} \right)+ {\cal O}(\lambda^3)~.
\ee
%%%%%%%%%%%%%%%%%%%%%%%%%%%%%%%%%%%%%%%%
%
The Majorana counterparts to $J_{CP}$, $S_1$ and $S_2$, are
given by 
%%%%%%%%%%%%%%%%%%%%%%%%%%%%%%%%%%%%%%%%%
\bea
S_1 \simeq  -\frac{\D \lambda}{\D 2}  \,  
\left( s_{\omega + \sigma} - B \, s_{\phi - \psi + \sigma} 
\right) 
\\[0.3cm]
S_2 \simeq -\frac{\D \lambda}{\D 2}  \, 
\left( s_{\omega - \rho + \sigma} - B \, s_{\phi - \psi - \rho + \sigma} 
\right) ~.
\eea
%%%%%%%%%%%%%%%%%%%%%%%%%%%%%%%%%%%
%
For $\rho = \sigma = 0$ one finds that to order $\lambda$ 
the relation $S_1 = S_2 = -2\sqrt{2}~J_{CP}$ is valid.   
Relatively small $\omega$ is favored 
in this case. Furthermore, atmospheric neutrino 
mixing is predicted to be close to maximal, 
to be more precise, the lower limit 
$\sa \gs 0.97$ holds.

%%%%%%%%%%%%%%%%%%%%%%%%%%%%%%%%%%%%%%%%%%%%%%

\subsection{\label{sec:lala}Large $\lambda_{ij}$}

%%%%%%%%%%%%%%%%%%%%%%%%%%%%%%%%%%%%

\hskip 0.8cm We consider $\lambda_{ij}\geq 1/\sqrt{2}$
and choose again as small 
expansion parameters the $\epsilon_{ij}$ 
introduced in Eq.\ (\ref{eq:eij}).
For the oscillation observables one finds 
%%%%%%%%%%%%%%%%%%%%%%%%%%%%%%%%%%%%%%%%%
\be 
\ba 
\tan^2 \theta_{\rm sol} \simeq 1 - 4 \, 
\frac{\D \cos (\omega + \phi)/2}{\D \cos (\omega - \phi)/2}  \, \epsilon_{12} 
+ 2 \, \frac{\D 2 + 2 \, c_{\omega + \phi} + c_{\omega + \psi} - 
c_{\phi + \psi}}{\D \cos^2 (\omega - \phi)/2}  \, \epsilon_{12}^2~,\\[0.3cm]
U_{e3} \simeq \sin (\omega - \phi)/2 + \epsilon_{12} \, 
\cos (\omega - \phi)/2~,
 \\[0.3cm]
\sin^2 2 \theta_{\rm atm}  \simeq  1 + {\cal{O}} (\epsilon^4)~.
\ea 
\ee
%%%%%%%%%%%%%%%%%%%%%%%%%%%%%%%%%%%%%%%%%%%%%%%
%
The atmospheric neutrino mixing is again
very close to maximal. The term 
$\sim \sin(\omega - \phi)/2$
in $U_{e3}$ is not suppressed by  
positive powers of $\epsilon_{ij}$. 
This suggests that $(\omega - \phi)/2$ should be 
relatively small. 
For $\omega = \phi$ the
expressions for the 
three rephasing invariants, 
associated with CP--nonconservation, take 
a rather simple form: 
%%%%%%%%%%%%%%%%%%%%%%%%%%%%%%%%%%%%%%%%%
\be
J_{CP} \simeq - \frac{s_\phi}{4} \, \epsilon_{12} + 
\frac{s_{\phi - \psi}}{4} \, \epsilon_{23}^2 ~,
\ee
%%%%%%%%%%%%%%%%%%%%%%%%%%%%%%%%%%%%%%%%%%%
%
and 
%%%%%%%%%%%%%%%%%%%%%%%%%%%%%%%%%%%%%
\bea
S_1 \simeq -\frac{\D s_{\psi - \sigma}}{\D \sqrt{2}} \, \epsilon_{12} - 
\frac{\D s_{\sigma}}{\D \sqrt{2}} \, \epsilon_{23}^2 ~,\\[0.3cm]
S_2 \simeq  \frac{\D s_{\psi + \rho - \sigma}}{\D \sqrt{2}} \, \epsilon_{12} - 
\frac{\D s_{\rho - \sigma}}{\D \sqrt{2}} \, \epsilon_{23}^2 ~.
\eea
%%%%%%%%%%%%%%%%%%%%%%%%%%%%%%%%%%%%%%%%
%
In the case of $\sigma = \rho = 0$ we have $S_1 \simeq - S_2$, 
but there is no simple relation between $S_{1,2}$
and $J_{CP}$.

Note that in all three cases under study the connection between 
$S_1$, $S_2$ and $J_{CP}$ is different. 

%%%%%%%%%%%%%%%%%%%%%%%%%%%%%%%%%%%%%%%%%

\subsection{\label{sec:cpca}Special Cases} 

%%%%%%%%%%%%%%%%%%%%%%%%%%%%%%%%%%%%%%%%%

\hskip 0.8cm It is 
instructive to examine the interplay 
of parameters when some of the 
three phases in $\tilde{U}_{\rm lep}$ 
and $P_\nu$ are zero. 
It turns out that there are no 
drastic consequences in the cases of 
``small'' hierarchical $\lambda_{ij}$, 
of $\lambda_{23} = 1$, and 
when all $\lambda_{ij}$ are ``large''. 
Interesting correlations 
happen, however, in the case of 
``small'' $\lambda_{ij}$ when the 
three  $\lambda_{ij}$
are of the same order.
To illustrate this we choose again 
$\lambda_{12} = \lambda$, $\lambda_{23} = A \, \lambda$ and 
$\lambda_{13} = B \, \lambda$ with 
$A, B$ real and of order one. 
We fix $\lambda = \sqrt{m_\mu/m_{\tau}} \simeq 0.24$ 
and vary $A, B$ between 0.4 and $1/\sqrt{\lambda}$. 
In doing so we observe that 
values of $A$ and $B$ slightly below 0.5 
are favored, indicating 
a  mild hierarchy in $\lambda_{ij}$. 
Inspecting Eq.\ (\ref{eq:mixcp}) one finds  
that the dependence on 
$\omega$ of the oscillation parameters 
is rather weak. If $\omega = \psi$ one finds from 
Eq.\ (\ref{eq:mixcp}) that \ts{} will be too large, $\ts \gs 0.58$.
We show in Fig.\ \ref{fig:cpsp} the scatter plots 
of the correlations between \ts{} and $|U_{e3}|^2$, 
and between 
$|U_{e3}|^2$ and $J_{CP}$,
which turn out to be the most interesting ones. 
The 1$\sigma$ allowed ranges of values of the 
neutrino mixing parameters given in 
Eq.\ (\ref{eq:range}) were used.
If the charged lepton mass Lagrangian 
conserves CP, i.e., if $\psi = 0$,
there exists a lower bound 
on $|U_{e3}|^2 \gs 0.005$. 
In this case $|J_{CP}|$ can be relatively large, 
for instance, $|J_{CP}| \simeq 0.035$
for $|U_{e3}|^2 = 0.025$, 
but is also allowed to be zero.

   If CP is conserved by 
the neutrino mass term, 
i.e., if $\phi = \omega = 0$, 
a simple correlation between 
\ts{} and $|U_{e3}|^2$ exists again:
smaller $|U_{e3}|^2$ implies smaller \ts{}
(see Fig.\ \ref{fig:cpsp}).
A similar limit on $|U_{e3}|^2$ as in
the case of $\psi = 0$
applies. Most remarkably,
the area of allowed values of $J_{CP}$ 
``bifurcates'' as a function of $|U_{e3}|^2$
and for values 
of $|U_{e3}|^2 \gs 0.008$,
$J_{CP} = 0$ is no longer 
allowed. For $|U_{e3}|^2 \geq 0.010$,
for instance, one has $|J_{CP}| \gs 0.01$. 
This can be understood as follows.
If $\omega = \phi = 0$ holds, the 
requirement of vanishing $J_{CP}$ 
(see Eq.\ (\ref{eq:JCPsmla})) 
is fulfilled for $\psi = 0$. Then, 
one sees from Eq.\ (\ref{eq:mixcp}) 
that to order $\lambda$, the equality 
%%%%%%%%%%%%%%%%%%%%%%%%%%%%%%%%%%%%%
\be \label{eq:cornohie}
\ts \simeq 1 - 4 \, \frac{1 - B}{1 + B} \, |U_{e3}| 
\ee
%%%%%%%%%%%%%%%%%%%%%%%%%%%%%%%%%%%%%%
%
holds. For the values of $B$ considered,
$B = (0.40 - 2.0)$, 
this equality is incompatible with 
the 1$\sigma$ upper bounds on 
\footnote{Note that, as is shown in Fig.\ \ref{fig:cpsp}, 
one has $J_{CP} = 0$ for the minimal allowed 
value of $|U_{e3}|^2 \simeq 0.006$. 
As can be seen from Eq.\ (\ref{eq:JCPsmla}),
the latter is reached in 
the case under discussion for the CP--conserving value
of $\psi = \pi$, for which $J_{CP}$ vanishes.} 
\ts{} and  $|U_{e3}|^2$.

   The ``bifurcation'' happens also when 
$\phi = \psi = 0$ holds (Fig.\ \ref{fig:cpsp}). 
The lower limit on $|U_{e3}|^2$ in this case 
is roughly by a factor of 4 smaller
than in the preceding ones. 
Though not shown here,
there can also be interesting 
constraints on \sa{} when some of the 
phases vanish. If, e.g., 
$\omega = \phi = 0$ holds, one sees 
from Eq.\ (\ref{eq:mixcp}) 
that $1 - \sa$ receives a correction 
of order $\lambda^2 \simeq 0.06$. 
It can be shown that typically $\sa \ls 0.98$ in this case. 

%%%%%%%%%%%%%%%%%%%%%%%%%%%%%%%%%%%%%%%

\subsection{\label{sec:spec}Specific Models}

%%%%%%%%%%%%%%%%%%%%%%%%%%%%%%%%%%%%%%

\hskip 0.8cm We shall consider next specific 
models in which Eq.\ (\ref{UPMNSCPV}) arises. 
We will also investigate how the CP--violating phases 
appear in the matrix $U_{\nu}$.
 
   It is well--known that when $U_{\rm lep} = {\mathbbm 1}$,
bimaximal neutrino mixing can 
be generated by the following 
neutrino mass matrix \cite{STPPD82,CLSP83}
%%%%%%%%%%%%%%%%%%%%%%%%%%%%%%%%%%%%%%
\be \label{eq:lelmlt}
m_\nu = \frac{m}{\sqrt{2}} \, 
\left( 
\bad 
0 & 1 & 1 \\[0.2cm]
1 & 0 & 0 \\[0.2cm]
1 & 0  & 0  
\ea 
\right)~.
\ee
%%%%%%%%%%%%%%%%%%%%%%%%%%%%%%%%%%%%%%
%
The latter has eigenvalues of 0 and $\pm m$. 
This matrix has a flavour symmetry which
corresponds to the 
conservations of the lepton charge
$L' = L_e - L_\mu - L_\tau$. 
One can lift the 
degeneracy of the two mass eigenvalues
by adding a small perturbation 
$\epsilon \, m/\sqrt{2}$ 
in the $\mu\tau$ entry.
This yields a mass matrix known 
from the Zee model \cite{Zee} (see also, e.g.,
\cite{FG99,Koide,FOY02,He}). 

    It is well--known that there is no 
CP--violation in the Zee model 
since one can absorb all three possible 
phases in the charged lepton fields. 
As we will see, this is true only when the charged 
lepton mass matrix is diagonal. 
One finds that in this case 
$\dma \simeq m^2 \, (1 - \epsilon/\sqrt{2})$ and 
$\dms \simeq \sqrt{2} \, m^2 \, \epsilon$. 
This fixes 
%%%%%%%%%%%%%%%%%%%%%%%%%%%%%%%%%%
\be 
\epsilon \simeq \frac{1}{\sqrt{2}} \, \frac{\dms}{\dma} 
\simeq \frac{1}{\sqrt{2}} \, \frac{7 \cdot 10^{-5} \, \rm eV^2} 
{2 \cdot 10^{-3} \, \rm eV^2} \simeq 0.025~,
\ee
%%%%%%%%%%%%%%%%%%%%%%%%%%%%%%%%%%
%
where we used the best--fit values found in the 
recent analyzes in \cite{BCGPRSNO3,SKatm03}.
For $\epsilon \neq 0$ the solar mixing is slightly 
reduced from its maximal value \cite{Koide,FOY02}: 
we have $\ts \simeq 1 - \epsilon/\sqrt{2} \simeq 0.98$.
This is incompatible even with the 5$\sigma$
allowed range of values of \ts{} (see, e.g., \cite{BCGPRSNO3}).
 
   In order to examine the implications
of existence of phases in $m_\nu$, 
we will assume that
the three non--vanishing entries in $m_\nu$ 
are complex and will denote the phases 
of these entries --- 
the $e \mu$, $e \tau$ and $\mu \tau$, 
by $\alpha'$, $\beta'$ 
and $\gamma'$, respectively. 
The neutrino mass Lagrangian then 
has the form:
%%%%%%%%%%%%%%%%%%%%%%%%%%%%%%%%%%%%%%
\be
{\cal{L}} = - \frac{m}{2\sqrt{2}}\left (
% \propto 
\overline{\nu_{e L}} \, \nu_{\mu R}^c \, e^{-i \alpha'} 
+ \overline{\nu_{e L}} \, \nu_{\tau R}^c \, e^{-i \beta'} + 
 \epsilon \,\overline{\nu_{\mu L}} \, 
\nu_{\tau R}^c \, e^{-i \gamma'} \right ) 
+ {\rm h.c.} 
\label{eq:CPZee}
\ee
%%%%%%%%%%%%%%%%%%%%%%%%%%%%%%%%%%%%%%
%
where $\nu_{l R}^c \equiv C \, \overline{\nu_{e L}}^{\rm T}$,
$C$ being the charge conjugation matrix.
We can redefine the flavor neutrino fields as follows: 
$\nu_{lL} = e^{i a_l} \nu'_{lL}$, $l = e, \mu, \tau$. 
Choosing 
%%%%%%%%%%%%%%%%%%%%%%%%%%%%%%%%%
\bea
a_e = - \frac{1}{2} (\alpha' + \beta' - \gamma')~, \\[0.2cm]
a_\mu = - \frac{1}{2} (\alpha' - \beta' + \gamma')~, \\[0.2cm]
a_\tau = - \frac{1}{2} (\beta' - \alpha' + \gamma')~,
\eea
%%%%%%%%%%%%%%%%%%%%%%%%%%%%%%%%%%%%%
%
we get in terms of the fields $\nu'_{lL}$:
%%%%%%%%%%%%%%%%%%%%%%%%%%%%%%%%%%%%%%
\be
{\cal{L}} = - \frac{m}{2\sqrt{2}}\left (
% \propto 
\overline{\nu'_{eL}} \, \nu_{\mu R}^{' c} 
+ \overline{\nu'_{eL}} \, \nu_{\tau R}^{'c}  + 
 \epsilon \, \overline{\nu'_{\mu L}} \, \nu_{\tau R}^{'c} \, \right )  
+ {\rm h.c.} 
\ee
%%%%%%%%%%%%%%%%%%%%%%%%%%%%%%%%%%%%%%
%
Now the neutrino mass 
matrix contains only real entries
and is diagonalized by
an orthogonal transformation with an 
orthogonal real matrix $O$. 
The matrix $O$ has to a
good approximation 
the bimaximal mixing form, Eq.\ (\ref{eq:Ubimax}). 
The mass eigenstate Majorana fields
$\chi_j$, $j=1,2,3$,
are related to the fields
$\nu'_L \equiv (\nu'_{eL}, \nu'_{\mu L}, \nu'_{\tau L})^T$ 
via $\nu_L' = O \, \chi_L$,  where
$\chi_L \equiv (\chi_{1L}, \chi_{2L}, \chi_{3L})^T$.
Since $\nu'_L$ can be written as $\tilde{P}^{\dagger}_\nu \, \nu_L$, where 
$\nu_L \equiv (\nu_{e L}, \nu_{\mu L}, \nu_{\tau L})^T$ 
contains the original flavor neutrino fields, and 
$\tilde{P}_\nu \equiv {\rm diag}(e^{i a_e}, e^{i a_\mu}, e^{i a_\tau})$,  
we have $\nu_L = \tilde{P}_\nu \, O \, \chi_L$. 
We can factor out one of the three phases in $\tilde{P}_\nu$, i.e., 
%%%%%%%%%%%%%%%%%%%%%%%%%%%%%%%%%
\be
\tilde{P}_\nu = e^{- \frac{i}{2} (\alpha' + \beta' - \gamma')} \, 
{\rm diag}(1, e^{- i(\gamma' - \beta')}, e^{-i(\gamma' - \alpha')}) 
\equiv e^{i \phi_\nu} \, P_\nu~.
\ee
%%%%%%%%%%%%%%%%%%%%%%%%%%%%%%%%%%
%
Thus, comparing the resulting formula 
with Eq.\ (\ref{eq:pmnscpv}), 
we see that in this special 
case of only three phases in $m_\nu$, 
for the phase matrix $Q_\nu$ holds: $Q_\nu = {\mathbbm 1}$. 
Consequently, the characteristic leading order relations between 
$S_1$, $S_2$ and $J_{CP}$ as discussed in Sections \ref{sec:smla} 
to \ref{sec:lala} will hold  
in the three generic structures of $U_\lambda$ discussed above. 
In terms of the notations used earlier in 
this Section we have $\phi = (\beta' - \gamma')$ and  
$\omega = (\alpha' - \gamma')$. 
Thus, there are altogether 3 
CP--violating phases in 
$\tilde{U}_{\rm lep}$ and
$U_\nu$ in the specific model 
we are considering. 
In the case of $\alpha' = \beta'$, 
one has $\phi = \omega$. 

% \noindent 
 The mixing parameters \ts{}, $U_{e3}$ and \sa{}
are functions of the 3 angles and 
1 phase in $\tilde{U}_{\rm lep}$ and of the 
2  phases in $P_\nu$. 
The formulae and results we have 
derived in Subsections \ref{sec:smla} to \ref{sec:cpca}
are valid in the specific model
under discussion if one sets
$\rho = \sigma = 0$ and
takes into account the fact 
that $\phi = (\beta' - \gamma')$ and  
$\omega = (\alpha' - \gamma')$.
If $U_{\rm lep} = \mathbbm{1}$, the phases in $P_\nu$ 
are unphysical and can be absorbed by
the charged lepton fields, CP is 
conserved in the lepton sector and $J_{CP} = S_1 = S_2 = 0$.

   The neutrino mass term, Eq.\ (\ref{eq:CPZee}),
leads to a neutrino mass spectrum with inverted hierarchy:   
$m_2 \simeq - m_1$, $m_3 \ll m_2$, and 
$\dms = m^2_2 - m^2_1 > 0$, $\dma = m^2_2 - m^2_3 \simeq m^2 \gg \dms$. 
The effective Majorana mass  measured in
\betabeta--decay, \meff, is given approximately by 
(see, e.g., \cite{BPP1}) 
%%%%%%%%%%%%%%%%%%%%%%%%%%%%%%%%%%%%%%%%%
\be \label{eq:meffcpsp} 
\meff \simeq \sqrt{\dma} \, \left |U_{e 1}^2 -  U_{e 2}^2\right | 
% \propto 
\simeq
\left\{ 
\baz
\sqrt{2}~ m\,\lambda \, \left | 1 - B \, e^{i (\alpha' - \beta' - \psi)} 
\right |~, 
&  \mbox{case I}  \\[0.3cm]
\sqrt{2}~ m\,\lambda \, \left | 1 + B \, e^{i (\beta' - \alpha' - \psi)} 
\right |~, 
&  \mbox{case II} 
\ea 
\right.
\ee
%%%%%%%%%%%%%%%%%%%%%%%%%%%%%%%%%%%%%%%%%%
%
where case I and case II refer respectively to 
$\lambda_{12} = \lambda$, 
$\lambda_{23} = A \, \lambda$, $\lambda_{13} = B \, \lambda$, 
and to $\lambda_{12} = \lambda$, 
$\lambda_{23} = 1$, $\lambda_{13} = B \, \lambda$.
Note that \meff{} does not vanish in the limit of
$\epsilon = 0$ \cite{STPPD82}.
Comparing the expressions for \meff{} with that for $J_{CP}$,
%%%%%%%%%%%%%%%%%%%%%%%%%%%%%%%%%%%%%%%%%%%%%%%%%%%%%
\be
J_{CP} \simeq  \left\{ 
\baz
\frac{1}{4 \sqrt{2}} (s_{\beta' - \gamma'} + 
 B \, s_{\alpha' - \gamma' - \psi}) \, \lambda~, 
 &  \mbox{case I}~, \\[0.3cm]
 \frac{1}{4 \sqrt{2}} \, (s_{\alpha' - \gamma'} - 
 B \, s_{\beta' - \gamma' - \psi}) \, \lambda ~,
&  \mbox{ case II}~,
 %\\[0.3cm]
% \frac{1}{4 \sqrt{2}} \, (s_{2\beta - \alpha - \gamma - 2 \psi} 
% + A \, s_{2 \beta - \alpha - \gamma - \psi} - B \, s_{\alpha - \gamma}) 
% \, \lambda^2 
% %& \mbox{ for case III}  
\ea 
\right. 
\ee
%%%%%%%%%%%%%%%%%%%%%%%%%%%%%%%%%%%%%%%%%%%%
%
one finds that all three phases 
$(\alpha' - \gamma')$, $(\beta' - \gamma')$ (or $\phi$, $\omega$)
and $\psi$ contribute to both, 
the ``Dirac'' phase $\delta$ and to 
the single Majorana phase that 
enters into the expression for \meff{} in the case of 
neutrino mass spectrum with inverted hierarchy \cite{BGKP96,0vbbCP}. 
 
    If we set $\beta' = \alpha'$, only $\psi$ contributes to \meff, while
$J_{CP}$ is a non--trivial function of the two phases
$(\alpha' - \gamma') \equiv \phi$ and $(\alpha' - \gamma' - \psi) 
\equiv (\phi - \psi)$. If, however, $\beta' = \gamma'$, then 
$\phi = 0$, $\omega = (\alpha' - \beta')$,
and we have to leading order in $\lambda$ 
in case I: $J_{CP} \simeq \lambda \, B \, 
s_{\alpha' - \beta' - \psi}/(4 \sqrt{2}) 
\simeq S_1/(2\sqrt{2}) \simeq S_2/(2\sqrt{2})$. 
%, i.e., Eq. (\ref{eq:JS12}) is valid. 
The effective Majorana mass as given in Eq.\ (\ref{eq:meffcpsp}) 
can be written as:
%%%%%%%%%%%%%%%%%%%%%%%%%%%%%%%%%%%%%%%%%
\be
\meff \simeq  m\, \left |\cos2\theta_{\rm sol} - i\,8 \, J_{CP} \right |~. 
\label{eq:meffJCP}
\ee
%%%%%%%%%%%%%%%%%%%%%%%%%%%%%%%%%%%%%%%%%%
%
The same relation holds in case II when $\alpha' = \gamma'$.  
Thus, quite remarkably, in these cases 
the deviation of \meff{} from the minimal value it can have 
in the case of neutrino mass spectrum with inverted
hierarchy (see, e.g., \cite{BPP1}), 
${\rm min}(\meff) \simeq m\cos2\theta_{\rm sol}$,
is determined by $J_{CP}$.

  A similar analysis can be performed and analogous 
results can be obtained if
the small term $\sim \epsilon$,
generating the requisite 
splitting between the two non--zero
eigenvalues of the matrix (\ref{eq:lelmlt}),
is present in one of the diagonal entries in Eq.\ (\ref{eq:lelmlt}). 

%%%%%%%%%%%%%%%%%%%%%%%%%%%%%%%%%%%%%%%%%%%%%%%

\section{\label{sec:concl}Conclusions}

%%%%%%%%%%%%%%%%%%%%%%%%%%%%%%%%%%%%%%%%%
%
\hskip 0.8cm Assuming that the three flavour neutrino mixing matrix 
$U_{\rm PMNS}$ a product 
of two unitary matrices
$U_{\rm lep}$ and $U_{\nu}$
arising from the diagonalization of 
the charged lepton and neutrino mass matrices,
$U_{\rm PMNS} = U^{\dagger}_{\rm lep} U_{\nu}$,
and that apart from possible CP--violating
phases, $U_{\nu}$ has a {\it bimaximal mixing
form}, Eq.\ ({\ref{eq:Ubimax}),
we have performed in the present article
a systematic study of the possible
forms of the matrix $U^{\dagger}_{\rm lep}$
which are compatible with the existing data 
on neutrino mixing and oscillations.
The three mixing angles in 
$U_{\rm lep}$ were treated 
as free parameters without assuming 
any hierarchy relation between them.
The case of CP--nonconservation
was of primary interest and 
was analyzed in detail.

   We have found that there exist
three possible generic structures 
of $U_{\rm lep}$, 
which are compatible with 
the existing data on neutrino mixing.
One corresponds to a hierarchical
``CKM--like'' matrix. In this case 
relatively large values of 
the solar neutrino mixing angle
$\theta_{\rm sol}$,  
and of $|U_{e3}|^2 \equiv |(U_{\rm PMNS})_{e3}|^2$,
are typically predicted, 
$\tan^2\theta_{\rm sol} \gtap 0.42$,
$|U_{e3}|^2 \gtap 0.02$,
while the atmospheric
neutrino mixing angle $\theta_{\rm atm}$ can
deviate noticeably from $\pi/4$,
$\sin^22\theta_{\rm atm} \gtap 0.95$. 
The second corresponds to
{\it one of the mixing angles in $U_{\rm lep}$
being equal to $\pi/2$}, 
and predicts practically maximal atmospheric 
neutrino mixing
$\sin^2 2 \theta_{\rm atm} \simeq 1$.
Large atmospheric neutrino 
mixing, $\sin^22\theta_{\rm atm} \gtap 0.95$,
is naturally predicted by the third possible 
generic structure of $U_{\rm lep}$, 
which corresponds to {\it all three
mixing angles in $U_{\rm lep}$ being large}.

  The parametrization of the
3 flavour neutrino mixing matrix as
$U_{\rm PMNS} = U^{\dagger}_{\rm lep} U_{\nu}$
is useful especially for studying the case 
of CP--nonconservation. We have found that, in general,
in addition to the three Euler--like angles,
$U^{\dagger}_{\rm lep}$ contains only one
CP--violating phase,
while  $U_{\nu}$ includes four CP--violating
phases and can be parametrized as follows:
$U_{\nu} = P_{\nu}~U_{\rm bimax}~Q_{\nu}$,
where  $P_{\nu}$ and $Q_{\nu}$ are diagonal phase 
matrices containing two phases each, and 
$U_{\rm bimax}$ is the bimaximal mixing matrix, 
Eq.\ ({\ref{eq:Ubimax}).
The two phases in $Q_{\nu}$
are Majorana--like --- the
flavour neutrino oscillation 
probabilities do not depend on them. In general, 
the three angles and the Dirac CP--violating phase %in \pmns{} 
in the PMNS matrix \pmns{} depend in a complicated manner
on the three angles and the phase in 
$U_{\rm lep}$ and on the two phases in $P_\nu$. % and $Q_{\nu}$.
Finally, the two Majorana CP--violating phases in
\pmns{} are functions of the
all the five CP--violating phases in
$U_{\rm lep}$, $P_\nu$ and
$Q_\nu$, and of the the three angles in
$U_{\rm lep}$. 

    We have derived approximate 
and rather simple expressions 
for the observables
\ts{}, $|U_{e3}|$ and \sa, as well as
for three rephasing invariants
of the \pmns{} matrix, $J_{CP}$ and $S_{1,2}$,
associated respectively with
CP--violation due to the Dirac phase
and due to the two Majorana phases
in \pmns{}. This is done for
each of the three possible 
generic structures of $U_{\rm lep}$. 
In certain specific cases we find that 
simple relations between $J_{CP}$ and $S_{1,2}$ hold
(see, e.g., Eq.\ (\ref{eq:JS12})).

    Finally, we considered a simple 
model in which 
$U_{\rm PMNS}$ is expressed in terms 
$U^{\dagger}_{\rm lep} U_{\nu}$
with $U_{\nu} = P_{\nu}~U_{\rm bimax}~Q_{\nu}$.
In this model the neutrino mass
term is assumed to have the same form
(when we set $U_{\rm lep} = \mathbbm{1}$)
as that of the Zee model.   
One finds that in this case $Q_\nu = \mathbbm{1}$,
while $P_{\nu}$ contains two CP--violating
phases. Expressions for the 
earlier indicated neutrino mixing observables
and for the effective Majorana mass in \betabeta--decay,
\meff, are given. We find that if the 
CP--violating phases in $P_{\nu}$
satisfy a certain condition, \meff{} depends in a 
simple way on 
$J_{CP}$ (Eq.\ (\ref{eq:meffJCP})). 

It would be interesting to generalize to the matrix 
$U_{\rm lep}~P_{\nu}$ the idea of unitarity triangles,
which has been successfully used in analyzing 
CP--violation in the quark sector, particularly
in $B$--meson decays. This could provide a
helpful graphical representation  
of CP--violation in neutrino oscillations.

\vspace{0.5cm}
\begin{center}
{\bf Acknowledgments}
\end{center} 
It is a pleasure to thank S.~Choubey and A.~Smirnov for helpful discussions. 
W.R. would like to acknowledge with gratefulness the hospitality 
of the Physics Department at Dortmund University, where 
part of this study was done.   
This work was supported in part by the EC network HPRN-CT-2000-00152 (W.R.), 
by US Department of Energy Grant DE-FG02-97ER-41036 (P.H.F.), 
and by the Italian INFN under the program ``Fisica Astroparticellare''
(S.T.P.).

\pagestyle{empty}
\begin{figure}[p]
\begin{center}
\hspace{-3cm}%\vspace{-3cm}
\epsfig{file=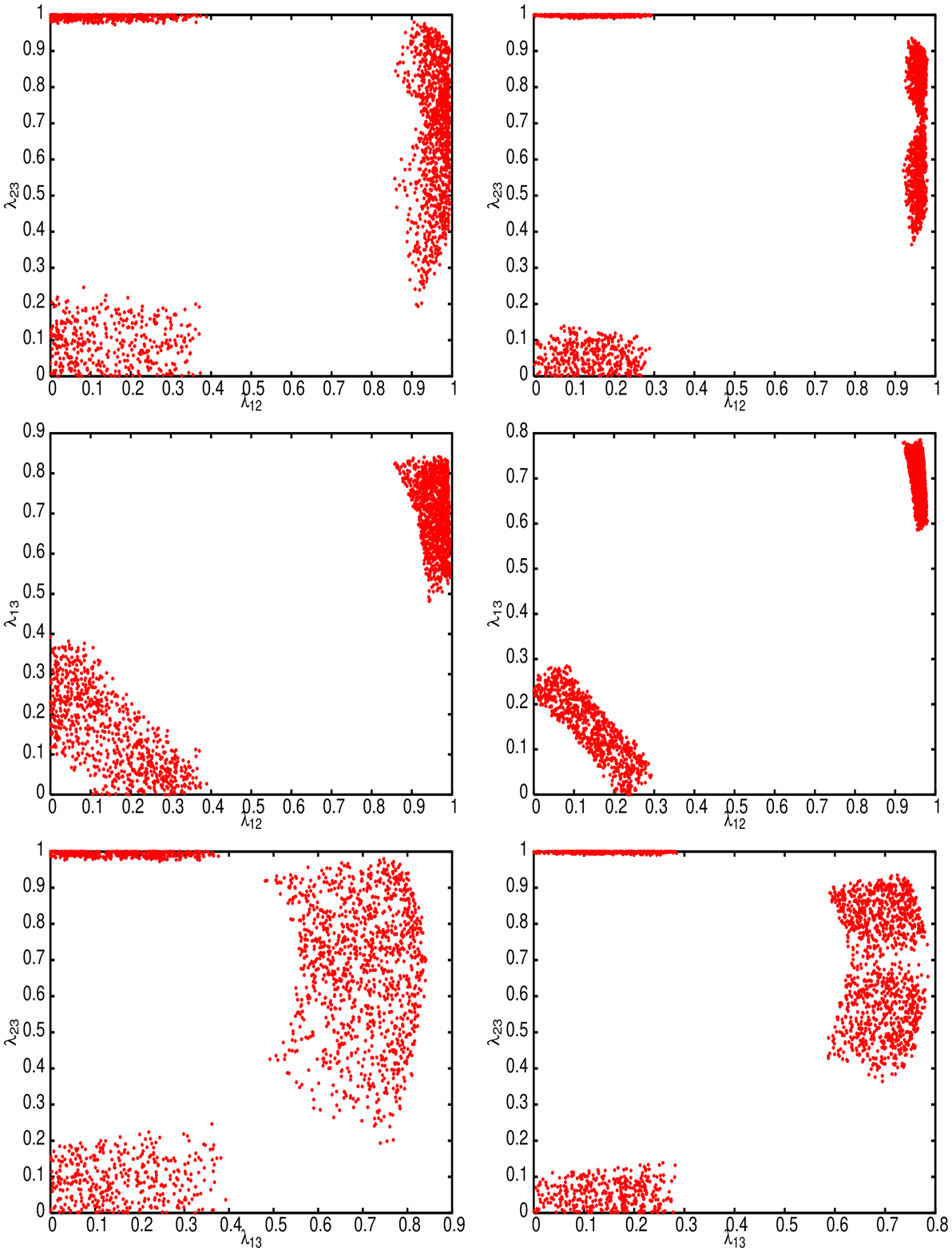,width=19cm,height=24cm}
\caption{\label{fig:mostgencpc}
Scatter plot of the three $\lambda$ parameters 
for the 3$\sigma$ (left column) and 1$\sigma$ 
(right column) allowed ranges of values of the 
neutrino mixing parameters given 
in Eq.\ (\ref{eq:range}).
Conservation of CP is assumed.}
\end{center}
\end{figure}

\pagestyle{empty}
\begin{figure}[p]
\begin{center}
\hspace{-3cm}%\vspace{-3cm}
\epsfig{file=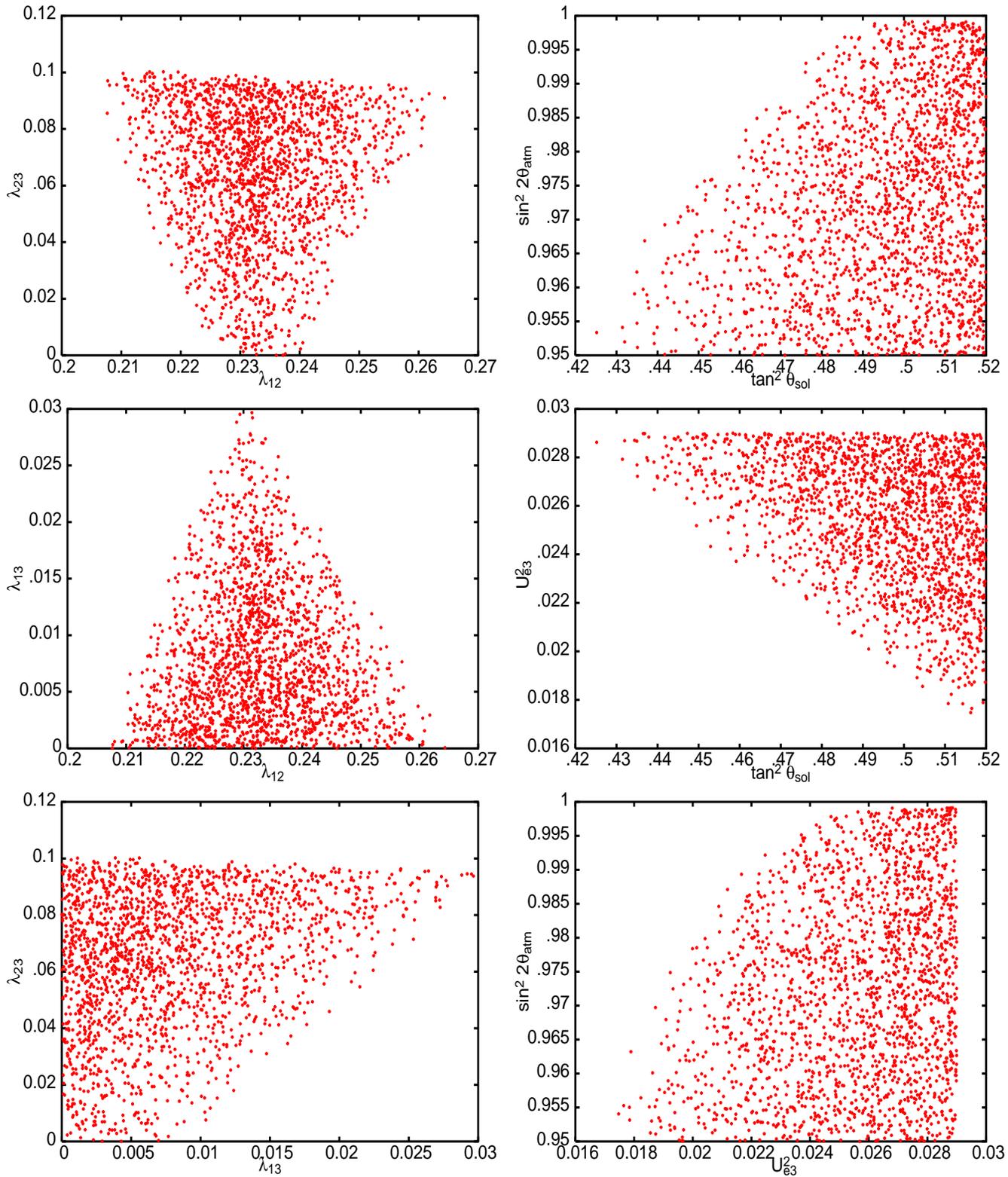,width=19cm,height=24cm}
\caption{\label{fig:cpcsmla}
Scatter plot of the $\lambda$  
and neutrino oscillation 
parameters in the case of ``small'' $\lambda_{ij}$ 
for the 1$\sigma$ allowed ranges of 
values of the neutrino mixing parameters 
given in Eq.\ (\ref{eq:range}).
Conservation of CP is assumed. }
\end{center}
\end{figure}

\pagestyle{empty}
\begin{figure}[p]
\begin{center}
\hspace{-3cm}%\vspace{-3cm}
\epsfig{file=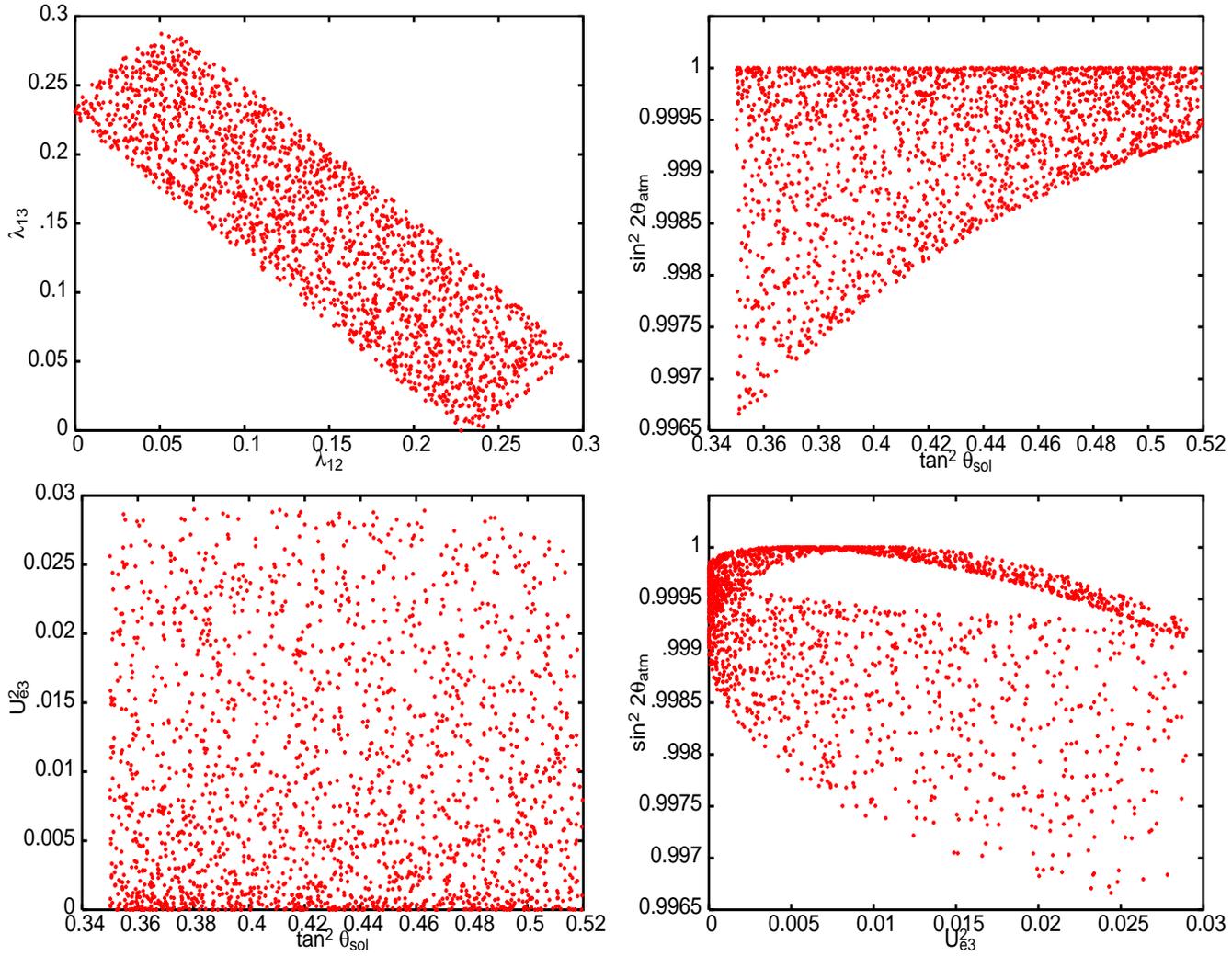,width=19cm,height=24cm}
\vspace{-5cm}
\caption{\label{fig:cpclam231}
Scatter plot of the two $\lambda$ parameters 
and the oscillation parameters in the 
case of $\lambda_{23} = 1$  
for the 1$\sigma$ range of the 
neutrino mixing parameters given 
in Eq.\ (\ref{eq:range}). 
Conservation of CP is assumed. }
\end{center}
\end{figure}

\pagestyle{empty}
\begin{figure}[p]
\begin{center}
\hspace{-3cm}%\vspace{-3cm}
\epsfig{file=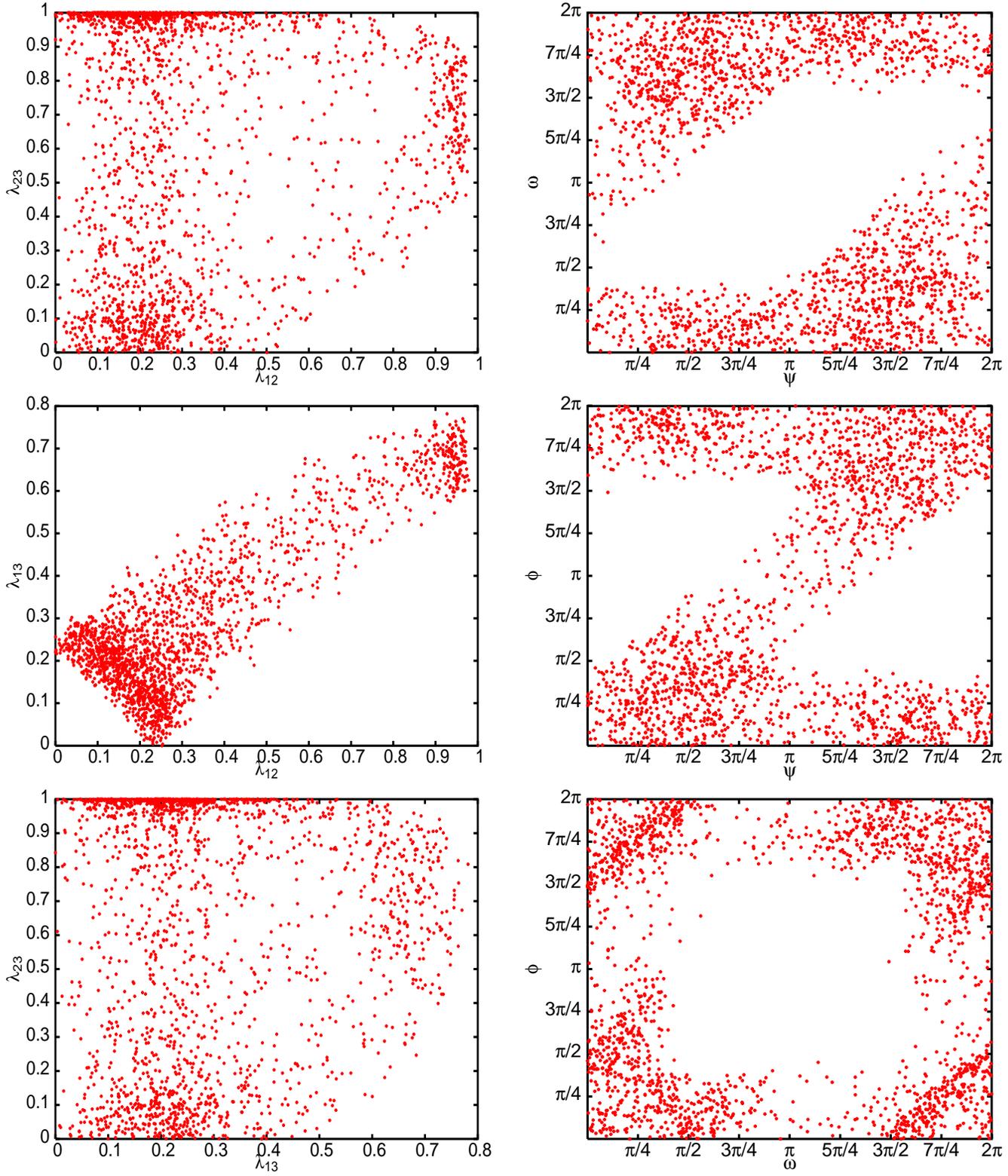,width=19cm,height=24cm}
\caption{\label{fig:cpv}
Scatter plot of the three $\lambda$ parameters 
and the CP--violating phases 
for the 1$\sigma$ allowed ranges of 
values of the neutrino 
mixing parameters given in Eq.\ (\ref{eq:range}).}
\end{center}
\end{figure}

\pagestyle{empty}
\begin{figure}[p]
\begin{center}
\hspace{-3cm}%\vspace{-3cm}
\epsfig{file=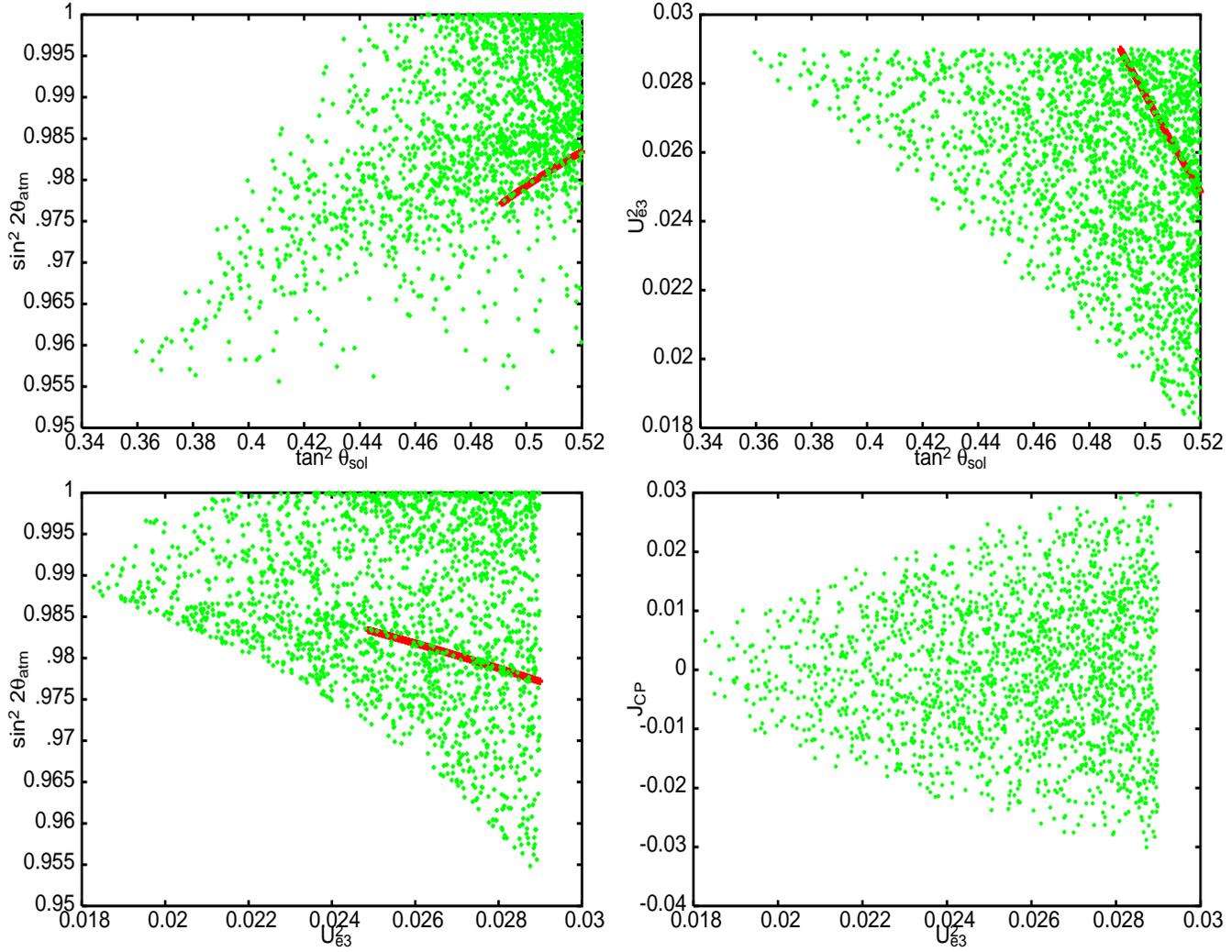,width=19cm,height=24cm}
\vspace{-5cm}
\caption{\label{fig:hie}Scatter plot of 
the oscillation parameters 
for the case of hierarchical $\lambda_{ij}$, i.e., 
$\lambda_{13} = \lambda_{12} \, \lambda_{23} = \lambda_{12}^3$. 
The various possible regions 
and correlations are shown both in the case of
CP--violation (bright green areas) and 
CP--conservation (dark red areas). 
The 1$\sigma$ allowed ranges of 
values of the neutrino mixing parameters 
given in Eq.\ (\ref{eq:range}) were used.}
\end{center}
\end{figure}

\pagestyle{empty}
\begin{figure}[p]
\begin{center}
\hspace{-3cm}%\vspace{-3cm}
\epsfig{file=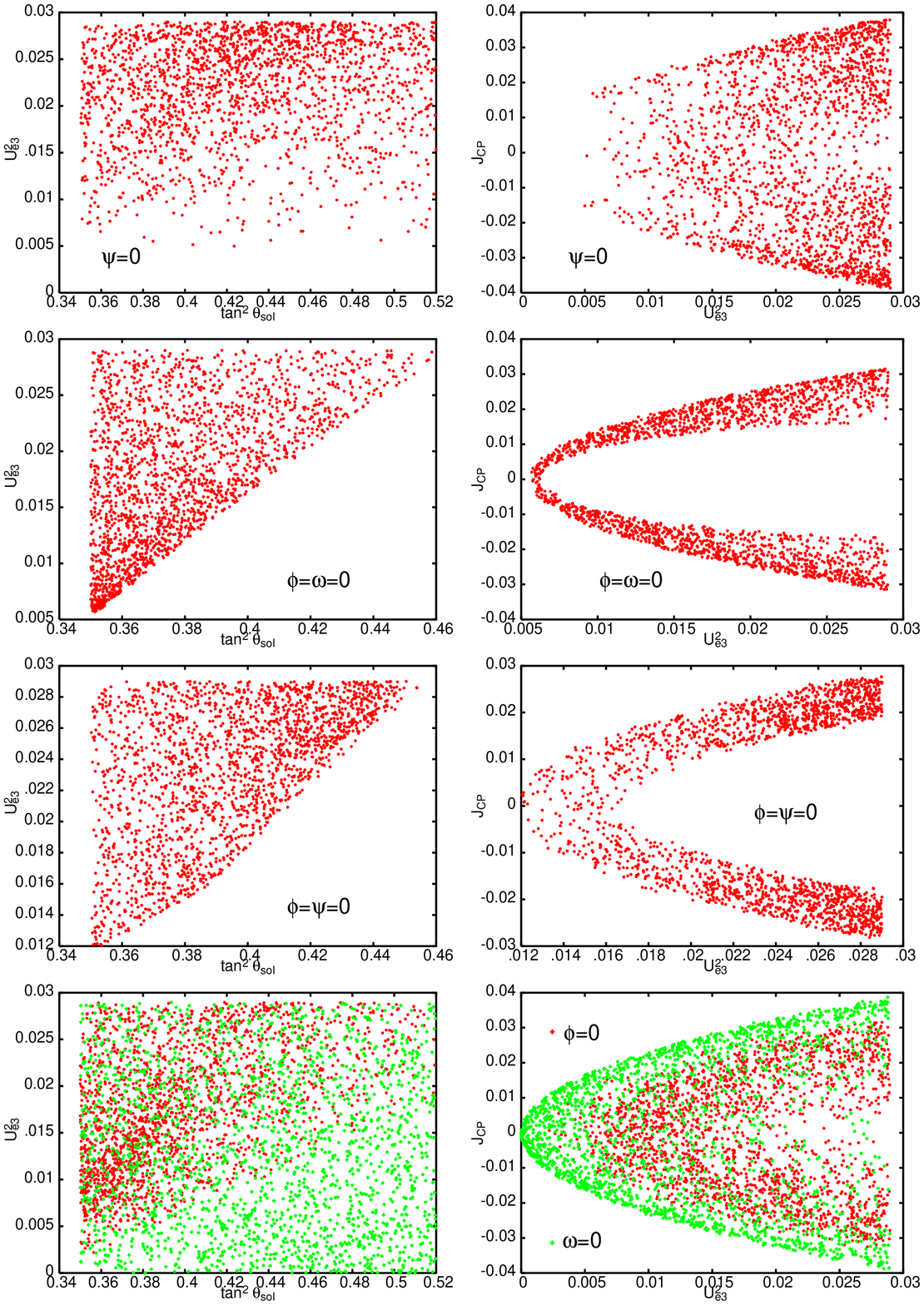,width=19cm,height=24cm}
\caption{\label{fig:cpsp}Scatter plot of 
the correlations between \ts{} 
% against 
and $|U_{e3}|^2$, and between $J_{CP}$ and $|U_{e3}|^2$, 
% against  
when the $\lambda_{ij}$ are of the 
same order and some of the 
CP--violating phases, 
$\psi$, $\phi$ and $\omega$, are zero. 
In the last row we display 
the CP--violating cases corresponding to 
$\omega = 0$ (bright green areas) and $\phi = 0$ 
(dark red area). 
The value $\lambda_{12} = 0.24$ 
% was chosen and the 
and 1$\sigma$ allowed ranges of values of the 
neutrino mixing parameters given in 
Eq.\ (\ref{eq:range}) were used.  }
\end{center}
\end{figure}

\end{document}